\documentclass[aps,prd,showpacs,superscriptaddress]{revtex4}

\usepackage{amsfonts}

\usepackage{amsmath}
\usepackage[USenglish]{babel}
\usepackage{graphicx}
\usepackage{color}
\usepackage{float}
\usepackage{multirow}

\usepackage{soul}

\begin{document}

\title{Exact solution of the Schr\"{o}dinger equation for an hydrogen atom at the interface between the vacuum and a  topologically insulating surface}

\author{D. A. Bonilla}
\email{dabonilla@uc.cl}
\affiliation{Facultad de F\'{i}sica, Pontificia Universidad Cat\'{o}lica de Chile, Vicu\~{n}a Mackenna 4860, Santiago, Chile}

\author{A. Mart\'{i}n-Ruiz}
\email{alberto.martin@nucleares.unam.mx}
\affiliation{Instituto de Ciencia de Materiales de Madrid, CSIC, Cantoblanco, 28049 Madrid, Spain}
\affiliation{Centro de Ciencias de la Complejidad, Universidad Nacional Aut\'{o}noma de M\'{e}xico, 04510 M\'{e}xico, Distrito Federal, M\'{e}xico}
\affiliation{Instituto de Ciencias Nucleares, Universidad Nacional Aut\'{o}noma de M\'{e}xico, 04510 M\'{e}xico, Distrito Federal, M\'{e}xico}

\author{L. F. Urrutia}
\email{urrutia@nucleares.unam.mx}
\affiliation{Instituto de Ciencias Nucleares, Universidad Nacional Aut\'{o}noma de M\'{e}xico, 04510 M\'{e}xico, Distrito Federal, M\'{e}xico}

\date{\today}

\begin{abstract}
When an hydrogen atom is brought near to the interface between $\theta$-media, the quantum-mechanical motion of the electron will be affected by the electromagnetic interaction between the atomic charges and the $\theta$-interface, which is described by  an axionic  extension of  Maxwell electrodynamics in the presence of a boundary. In this paper we investigate the atom-surface interaction effects upon the  energy levels and wave functions of an hydrogen atom placed at the interface between a $\theta$-medium and the vacuum. In the approximation considered, the Schr\"{o}dinger equation can be exactly solved by separation of variables in terms of hypergeometic functions for the angular part and hydrogenic functions for the radial part. In order to make such effects apparent we deal with unrealistic high values of the $\theta$-parameter.  We also compute the energy shifts using perturbation theory for a particular small value of $\theta$ and we demonstrate that they are in a very good agreement with the ones obtained from the exact solution.
\end{abstract}


\maketitle

\section{Introduction}

In electrodynamics there is the possibility of writing two quadratic gauge and Lorentz invariant terms: the first one is the usual electromagnetic density $\mathcal{L} _{\rm{EM}} = (\textbf{E} ^{2} - \textbf{B} ^{2})/ 8 \pi$ which yields Maxwell's equations, and the second one is the magneto-electric term $\mathcal{L} _{\theta} = \theta \,  \textbf{E} \cdot \textbf{B}$, where $\theta$ is a coupling field usually known as the axion angle. Many of the interesting properties of the latter term can be recognized from its four dimensional structure $\mathcal{L} _{\theta} = - ( \theta / 4 ) \epsilon^{\mu\nu\rho\lambda} F_{\mu\nu} F_{\rho \lambda}$, where $\epsilon^{\mu\nu\rho\lambda}$ is the Levi-Civit\`{a} symbol and $F_{\mu\nu}$ is the electromagnetic field strength. When $\theta$ is globally constant, the term ${\cal L}_\theta$ is a total derivative and has no effect on Maxwell's equations. These properties qualify ${\cal P} = \epsilon ^{\mu\nu\rho\lambda} F_{\mu\nu} F_{\rho \lambda}$ to be a topological invariant. Actually, ${\cal P}$ is the simplest example of a Pontryagin density  \cite{PONT}, corresponding to the abelian group $U(1)$. This structure together with  its generalization to nonabelian groups, has been relevant in diverse topics in high energy physics such as anomalies \cite{ANOM}, the strong CP problem  \cite{SCP}, topological field theories  \cite{TFT} and axions  \cite{AXIONS}, for example. 

Recently, an additional application of the theory defined by the full  Lagragian density $\mathcal{L} _{\rm{EM}} -(\alpha/4 \pi^2) \mathcal{L} _{\theta} $,  which we call $\theta$-electrodynamics ($\theta$ED), has been highlighted in condensed matter physics, where a piecewise constant axion field $\theta$ provides an effective field theory describing the electromagnetic response of a topological insulator (TI) ($\theta = \pi$) in contact with a trivial insulator ($\theta = 0$) \cite{TI,TI2,TI3,TI4}. A constant $\theta$, called the magnetoelectric polarization (MEP) of the material, defines what we call a $\theta$-medium and can be considered as an additional parameter characterizing the material in a way analogous to the dielectric permittivity $\varepsilon$ and the magnetic permeability $\mu$, which nevertheless manifest only in the presence of a boundary where its value suddenly changes. 
Topological insulators in (3+1)D have attracted great attention in condensed matter physics. These materials display nontrivial topological order and are characterized by a fully insulating bulk together with gapless surface states, which are protected by time-reversal (TR) symmetry \cite{TI2, TI3}. This type of topological behavior was first predicted in graphene \cite{Kane-Mele}. It was subsequently predicted and then observed in alloys and stoichiometric crystals that display strong enough spin-orbit coupling to induce band inversion, such as Bi${}_{1-x}$Sb${}_{x}$ \cite{Exp-TI1, Exp-TI2}, Bi${}_{2}$Se${}_{3}$, Bi${}_{2}$Te${}_{3}$, Sb${}_{2}$Te${}_{3}$ \cite{Exp-TI3, Exp-TI4} and TlBiSe${}_{2}$ \cite{Sato}. Based on subtle field theoretical calculations, it was lately proposed that $\theta$ED also describes the electromagnetic response of Weyl semimetals \cite{WS}, in which the broken time-reversal or parity symmetries allow for a space-time dependent  MEP. Recently, the study of topological insulating and Weyl semimetal phases either from a theoretical or an experimental perspective has been actively pursued \cite{TI1, WS1}.

 The main consequence of $\theta$ED, which is manifest  in TIs , is the topological  magnetolectric (TME) effect whereby an electric (magnetic) field can induce a magnetic (electric) field even in the static case, as the consequence of the field-dependent effective charge an current densities (being proportional to ${\mathbf B}$ and  ${\mathbf E}$, respectively) that  derives from the term ${\cal L}_\theta$.  
The simplest manifestation of the TME effect in TIs is the generation of a magnetic field when a point like electrical charge is located in front of a TI \cite{science}. Even more interestingly, this field can be interpreted as arising from image magnetic monopoles located at the corresponding sides of the interface. Nevertheless, as it is well known, image charges in electrodynamics are only mathematical constructions lacking of physical significance which might be useful to provide an intuitive understanding of some phenomena. $\theta$ED postulates $\nabla \cdot {\mathbf B}=0$  in such a way that real magnetic monopoles are not allowed in the theory, as emphasized once more in Ref. \cite{Flavio}. The physical origin of this magnetic field relies in the Hall currents at the interface which are induced by the TME effect. The topology inherent to the band structure of TIs is reflected at the macroscopic level in the 
the quantization condition $\theta= \pm(2m+1)\pi, m \in Z$, which allows to distinguish between TI's and  normal insulators by assigning  the value $\vartheta=0$ modulo $2 \pi$ to the latter.

 The detection of the quantized TME effect provides an experimental challenge that has been pursued in different settings, which we briefly review. For example, measurements with a magnetic force microscope have been proposed in order to detect the magnetic field due to the image monopole produced by a point charge in front of a planar TI \cite{science}.
Another proposal consists in observing the dependence upon the MEP of the dispersion angle in a Rayleigh scattering of electromagnetic radiation incident upon circular cylinders fabricated with TIs . This requires the measurement of the electric field  components of the scattered waves in the far field region at one or two scattering angles \cite{Rayleigh}. 
The effect of the MEP on Kerr and Faraday rotations have also been considered to measure the TME effect. The Kerr effect consists of the rotation of the polarization plane of the reflected wave with respect to the incident plane and the Faraday effect refers to the analogous rotation of the  polarization plane of the transmitted wave. These effects take place in optically active materials, when the presence of an external magnetic field provides some anisotropy in the permittivity tensor. These optical rotations  have  also been predicted  to occur at the interface of a trivial insulator and a TI \cite{TI1,FARAD_KERR}.  In Ref. \cite{Maciejko}, an optical  experiment is proposed to observe the quantization of the MEP by measuring the Kerr and Faraday rotation angles in light shinning at normal incidence  from the vacuum to a planar TI  supported by an additional substrate of  a normal insulator. Measurements of the quantized  Kerr and Faraday angles in the presence of  TIs are reported in Refs. \cite{OKADA,WU,FARADAY2}. 
The observation of a Faraday rotation equal to the fine structure constant is described  in Ref.  \cite{DZIOM} and constitutes   a direct consequence of the TME effect which also confirms  $\theta$ED as the effective theory yielding  the electromagnetic response of 3D TIs .
The experimental determination of the TME effect arising from TIs in (3+1)D has proved to be challenging mainly because the topological response  always coexist with the ordinary electromagnetic response.

Within the realm of atomic spectroscopy and because of the well-developed theory together with  a large tradition in high precision measurements \cite{Swanberg}, 
hydrogenlike ions
could provide an attractive test bed for studying the TME effect, since their atomic structure is sensitive to the image monopole magnetic fields.
 In general terms, the presence of an atom in front of a material body modifies its quantum properties, such as the magnitude of the energy levels and the decay rates of the excited states. The atom-surface interaction depends on the particular material under consideration. For example, for an atom close enough to a dielectric body, the atom-surface interaction can be modeled by the influence of the image electric charges induced by the real atomic charges on the quantum mechanical motion of the atomic electrons. For materials whose electromagnetic response is more complicated than those of dielectrics, the atom-surface interaction can be extremely difficult and analytically intractable. In fact, when an hydrogenlike ion is brought near to an exotic material like a TI  the electric charges of the ion will polarize the material according to the usual Maxwell response term; but they also will  magnetize the medium due to the TME effect. Therefore, the effects of the $\theta$-medium on the quantum-mechanical motion of the atomic electron will be described by the Schr\"{o}dinger equation in the presence of the additional aforementioned magnetic field besides the usual nucleus-electron Coulomb potential. This problem has been recently considered in Refs. \cite{EPL} and \cite{PRA}. In the former investigation the authors report a (numerical) classical treatment of this system while in the latter a perturbative analysis of the realistic problem (with nontrivial optical properties) is given. The latter work also suggests
that the case of circular Rydberg ions will be of relevance because they provide an enhancement  of the TME effect contribution with respect to the optical one.

In general this problem is a complex task, for which in this paper we consider the idealized case where the atomic nucleus of an hydrogen atom is located at the interface between the vacuum and the $\theta$-medium. 
 This  provides a  simplified setting which allows a qualitative understanding of the consequences of the MEP upon the energy spectrum and the energy eigenfunctions. 
 For definiteness, we consider the situation in which the MEP $\theta$ is piecewise constant, and we isolate the effects of the $\theta$-term by considering different values of $\theta$ on each  side of the interface $z=0$. Also we take each medium with neither dielectric nor magnetic properties, i.e. we take $\varepsilon = \mu = 1$ everywhere.  It is worth mentioning that the formulation of $\theta$ED pursued in this work can be considered as a particularly simple version of the so-called Janus field theories, which have been actively explored in the context of the AdS/CFT correspondence \cite{Janus1, Janus2}. Also, when locating the nucleus just at the interface $z=0$, the TEM effect  in $\theta$ED produces the magnetic field of a Pearl vortex \cite{Pearl,Flavio}. In this way, our work yields the  energy eigenvalues and eigenfunctions  of an electron moving in the presence of such a vortex.

The paper is organized as follows. Section \ref{thetamedium} contains the basics of $\theta$ED emphasizing the appearance of the TME effect generated by the additional axion coupling \cite{Wilczek}. In Section \ref{QMP} we derive the Hamiltonian which describes the interaction between an hydrogen atom in vacuum near a $\theta$-medium. Here we clarify the  difference  between the motion of an electron in a magnetic monopole field ($\nabla \cdot {\mathbf B}\neq 0$) and  the motion in the image magnetic monopole field ($\nabla \cdot {\mathbf B}=0$) arising in $\theta$ED. We tackle the Schr\"{o}dinger equation by separation of variables. The solution of the angular part is presented in Section \ref{ANG_EQ} paying attention to the quantization of a new quantum number $L$ which generalizes the orbital angular momentum label $l$. Section \ref{RAD_EQ} deals with the solution of the radial equation together with the determination of the energy eigenvalues. In Section \ref{PERT_CALC} we show that the energy shifts calculated using perturbation theory coincides with the ones obtained from the exact result for a weak $\theta$-medium. Finally we close with a summary and some comments in Section \ref{SUMM}.

\section{Electrodynamics of a $\theta$-medium}
\label{thetamedium}

Electromagnetic phenomena involving matter are described by the Maxwell's field equations,
\begin{align}
& \nabla \cdot \textbf{D} = 4 \pi  \rho , \quad \nabla \cdot \mathbf{B} = 0 , \quad \nabla \times \mathbf{E} +\frac{1}{c} \frac{ \partial \mathbf{B}}{\partial t} = 0 , \quad \nabla \times \mathbf{H} - \frac{1}{c} \frac{ \partial \mathbf{D}}{ \partial t} =\frac{4 \pi}{c} \mathbf{J} , \label{MAXMM}
\end{align}
together with constitutive relations giving the displacement $\textbf{D}$ and the magnetic field $\textbf{H}$ in terms of the electric $\textbf{E}$ and magnetic induction $\textbf{B}$ fields \cite{Jackson}. These depend on the nature of the material, and they are generally of the form $\mathbf{D} = \mathbf{D} (\mathbf{E} ,\mathbf{B})$ and $\mathbf{H} = \mathbf{H} (\mathbf{E} ,\mathbf{B})$. For instance, for linear media they are $\textbf{D} = \varepsilon \textbf{E}$ and $\textbf{H} = \textbf{B} / \mu$, where $\varepsilon$ is the dielectric permittivity and $\mu$ is the magnetic permeability. For isotropic materials $\varepsilon$ and $\mu$ are scalar objects, while for anisotropic materials they are tensorial in nature. For realistic materials, the constitutive relations are nonlinear ($\varepsilon$ and $\mu$ are functions of $\textbf{E}$ and $\textbf{B}$). 

In this paper we are concerned with a particular class of bianisotropic materials described by the following constitutive relations
\begin{align}
\textbf{D} = \varepsilon \textbf{E} - \frac{\theta \alpha}{\pi} \textbf{B} , \qquad \textbf{H} = \frac{1}{\mu} \textbf{B} + \frac{\theta \alpha}{\pi} \textbf{E} , \label{ec3}
\end{align}
where $\theta$ is an additional parameter called the magnetoelectric polarization of the medium and $\alpha \simeq 1/137$ is the fine structure constant. In this case, the inhomogeneous Maxwell's equations reads
\begin{align}
\nabla \cdot( \varepsilon \textbf{E} )= 4 \pi \rho + \frac{\alpha}{\pi} \nabla \theta \cdot \textbf{B} , \qquad \nabla \times \left( \textbf{B} / \mu \right) -\frac{1}{c}\frac{\partial (\varepsilon \textbf{E} )}{\partial t}= \frac{4 \pi}{c} \textbf{J} - \frac{\alpha}{\pi} \nabla \theta \times \textbf{E} - \frac{1}{c} \frac{\alpha}{\pi} \frac{\partial \theta}{\partial t} \textbf{B} .
\label{MAXMOD}
\end{align}
These field equations have a wide range of applications in physics. For example, they describe (i) the electrodynamics of metamaterials when $\theta$ is a purely complex function \cite{MetaMat}, (ii) the electromagnetic response of topological insulators when $\theta = \pi$ \cite{TI} and (iii) the electromagnetic response of Weyl semimetals for $\theta (\textbf{x} , t) = 2 \textbf{b} \cdot \textbf{x} - 2 \hbar b _{0} t$ \cite{WS}. In fact, the modified Maxwell's equations (\ref{MAXMOD}) can be derived from the usual electromagnetic action supplemented with the $U(1)$ Pontryagin term:
\begin{align}
S[ \Phi, \textbf{A} ] = \int dt \, d ^{3} \textbf{x} \left[ \frac{1}{8 \pi} \left( \varepsilon \textbf{E} ^2 - \frac{1}{\mu} \textbf{B} ^2 \right ) - \frac{\alpha}{4 \pi ^{2}} \theta(\textbf{x}) \, \textbf{E} \cdot \textbf{B} - \rho \Phi + \frac{1}{c} \textbf{J} \cdot \textbf{A} \right]. \label{action}
\end{align}
The electromagnetic fields $\textbf{E}$ and $\textbf{B}$ are related with the electromagnetic potentials $\Phi$ and $\textbf{A}$ as usual due to the gauge invariance of the action (\ref{action}). For definiteness, in this work we deal with a very idealized situation in which the medium has neither dielectric nor magnetic properties, i.e. we take $\varepsilon = 1$ and $\mu = 1$, and we thought $\theta$ as a phenomenological nondynamical quantity characterizing the medium. We will refer to these materials as $\theta$-media and the theory describing their electromagnetic response via the extended action (\ref{action}) will be called $\theta$-electrodynamics or simply $\theta$ED. Next we comment upon two important consequences of the modified Maxwell's equations (\ref{MAXMOD}):

(a) The inhomogeneous Maxwell equations acquire  additional charge and current responses given by 
\begin{align}
\rho _{\theta} = \frac{\alpha}{4 \pi ^{2}} \nabla \theta \cdot \textbf{B} , \qquad \textbf{J} _{\theta} = - \frac{c \alpha}{4 \pi ^{2}} \nabla \theta \times \textbf{E} - \frac{\alpha}{4 \pi ^{2}} \frac{\partial \theta}{\partial t} \textbf{B} . \label{DENEFF}
\end{align}
Current conservation ($\nabla \cdot \textbf{J} _{\theta} + \partial \rho _{\theta} / \partial t = 0$) can be directly verified. Note that these expressions depend only on the space-time gradients of the axion angle. This is because the $\theta$-term, $\mathcal{L} _{\theta}$, is a total derivative for $\theta =$const, and does not affect the equations of motion. Physically, the effective charge and current densities (\ref{DENEFF}) encode the magnetoelectric effect of the $\theta$-medium \cite{Landauelectro}, which can be defined as a magnetization induced by an electric field, or alternatively, a charge polarization induced by a magnetic field.

(b) As pointed out, a globally constant axion angle has no effect on Maxwell's equations even though the constitutive relations depend upon $\theta$. This is why in this paper we focus on the simplest situation in which the axion angle produces nontrivial effects on the electromagnetic responses: a piecewise constant function $\theta (\textbf{x})$ exhibiting plane symmetry. In this scenario, the axion angle can be written as 
\begin{align}
\theta(z) = \theta _{1} H(-z) + \theta _{2} H(z) ,
\label{PWC} 
\end{align}
where $H(z)$ is the Heaviside function: $H(z) = 0$ for $z < 0$ and $H (z) = 1$ for $z > 0$. In this case, Maxwell's equations hold in the bulk regions where $\theta$ is a constant; however, the gradient of the axion angle has support only at the $\theta$-interface
\begin{align}
\nabla \theta = ( \theta _{2} - \theta _{1}) \delta(z) \hat{\textbf{e}} _{z} , \label{Gradient}
\end{align}
where $\delta(z)$ is the Dirac delta function and $\hat{\textbf{e}} _{z}$ is the unit vector in the direction $z$. In this way, the dynamical modifications in Eqs. (\ref{MAXMOD}) arise only at the boundary, which is the only place where the effective sources (\ref{DENEFF}) are nonzero. In this sense, a $\theta$-medium has conducting properties at the boundary, even though its bulk behaves as a normal insulator. In fact, for the model under consideration, the additional charge and current densities (\ref{DENEFF}) become $\rho _{\theta} = \tilde{\theta} \, \delta (z) \, \hat{\textbf{e}} _{z} \cdot \textbf{B}$ and $\textbf{J} _{\theta} = - \tilde{\theta} \, \delta (z) \,  \hat{\textbf{e}} _{z} \times \textbf{E}$; and these yield the following boundary conditions in the vicinity of the interface between two $\theta$-media:
\begin{gather}
[\hat{\textbf{e}} _{z} \cdot \textbf{E} ] = \tilde{\theta} \, \hat{\textbf{e}} _{z} \cdot \textbf{B} (z = 0) \quad , \quad  [\hat{\textbf{e}} _{z} \cdot \textbf{B}] = 0 , \\ [\hat{\textbf{e}} _{z} \times \textbf{B}] = - \tilde{\theta} \, \hat{\textbf{e}} _{z} \times \textbf{B} (z = 0) \quad , \quad [\hat{\textbf{e}} _{z} \times \textbf{E}] = 0 , 
\end{gather}
where $\tilde{\theta} = \alpha(\theta _{2} - \theta _{1}) / \pi$. The notation is $[\textbf{V} _{i}] = \lim _{\epsilon \rightarrow 0 ^{+}} \textbf{V} _{i} (z = + \epsilon) - \textbf{V} _{i} (z = - \epsilon)$ for any vector $\textbf{V}$. Many interesting magnetoelectric effects at the interface between two $\theta$-media have been highlighted using different approaches. For example, (i) the appearance of mirror magnetic monopoles when a pointlike electric charge is brought near to the $\theta$-interface \cite{science, Alberto1}, (ii) nontrivial Faraday and Kerr rotations appear when electromagnetic waves propagates across the interface \cite{Obukhov} and (iii) the Casimir stress on a $\theta$-interface between two metallic plates \cite{EPL2}.

\section{The quantum mechanical problem}
\label{QMP}

For an atom embedded in a $\theta$-medium with no boundaries, Maxwell's equations hold and thus the solution of the Schr\"{o}dinger equation is the usual one. In other words, the system does not feel any effect from the presence of the $\theta$-medium. However, when an atom is close to the interface between two $\theta$-media,  Maxwell's equations acquire additional magnetoelectric terms which yields to new interactions between the atomic electron and the interface, besides the usual nucleus-electron electromagnetic interaction. This problem has been tackled recently from both a classical \cite{EPL} and a quantum perturbative \cite{PRA} perspectives, but an analytic treatment of the quantum-mechanical problem is still absent. Our main aim here is to fill in this gap. In this section we derive the Schr\"{o}dinger equation describing the motion of the electron when a hydrogen atom is located at the interface between two $\theta$-media, and subsequently we solve in detail the resulting angular and radial differential equations.

\subsection{The electromagnetic potentials}

We consider two $\theta$-media in contact, having the MEP  modeled by the function (\ref{PWC}). The left-half space ($z<0$) is occupied by a $\theta$-medium, whereas the right-half space ($z>0$) is the vacuum. Now, let us begin with an hydrogen atom in the vacuum region near the $\theta$-interface $z=0$, such that the nucleus is fixed at $\mathbf{b} = b \hat{\textbf{e}} _{z}$ . Furthermore, in this paper we study the effects produced by the proton's electromagnetic fields near the $\theta$-medium on the atomic electron, and we neglect those induced by the position of the electron. With the purpose of deriving the Hamiltonian describing the quantum-mechanical motion of the atomic electron, let us take advantage of a convenient way to  deal with the EM fields produced by arbitrary sources $J ^{\mu} = (\rho , \textbf{J})$ in the presence of a $\theta$-medium, which is the introduction of the matrix Green's function (GF) $G ^{\mu} _{\phantom{\mu} \nu} \left( \mathbf{r} , \mathbf{r} ^{\prime} \right)$, such that the EM potential $A ^{\mu} = (\Phi , \textbf{A})$ is given by
\begin{align}
A ^{\mu} \left( \mathbf{r} \right) = \int d ^{3} \mathbf{r} ^{\prime} \; G ^{\mu} _{\phantom{\mu} \nu} \left( \mathbf{r} , \mathbf{r} ^{\prime} \right) J ^{\nu} \left( \mathbf{r} ^{\prime} \right) .  \label{GreenMatrix}
\end{align}
Such GF's have been calculated for plane, 
cylindrical and spherical geometries in Refs. 
\cite{Alberto1,Alberto2,Alberto3}. Taking the 
source as the proton 
charge density, i.e. $J ^{0} (\textbf{r}) = e \delta (\textbf{r} - \textbf{b}) $ and $\textbf{J} = 0$, where $e>0$ is the magnitude of the proton charge, the resulting EM potentials can be obtained in a simple fashion. See for example Ref. \cite{Alberto1}. For latter convenience, it is useful to express such potentials in the coordinate system attached to the nucleus, where they become
\begin{align}
\Phi(\textbf{r}) &= \left\lbrace \begin{array}{ccc} \frac{e}{r} - \frac{\tilde{\theta} ^{2}}{4 + \tilde{\theta} ^{2}} \frac{e}{\sqrt{\rho ^{2} + (z + 2b) ^{2}}}&;& z > 0 \\[9pt] \frac{4}{4 + \tilde{\theta} ^{2}}\frac{e}{r} & ; & z < 0 \end{array} \right. , \label{Hpotescalar} \\ \textbf{A} (\textbf{r}) &= \frac{2 e \tilde{\theta}}{4 +
 \tilde{\theta} ^{2}} \frac{y \hat{\textbf{e}} _{x} - x \hat{\textbf{e}} _{y}}{\rho ^{2}} \left( 1 - \frac{2b + \vert z \vert}{\sqrt{\rho ^{2} + (z + 2b) ^{2}}} \right) . \label{HpotA}
\end{align}
Here $r = \vert \textbf{r} \vert = \sqrt{\rho ^{2} + z ^{2}}$ is the proton-electron distance, $\rho ^{2} = x ^{2} + y ^{2}$, and the $\theta$-interface is now located at $z = - b$. The vector potential naturally satisfies the Coulomb gauge in each sector of the model. The Schr\"{o}dinger equation can thus be written using the previously determined EM potentials via the usual minimal coupling in the Hamiltonian
\begin{align}
H = \frac{1}{2 \mu} \left( \textbf{p} + e \textbf{A} \right) ^{2} - e \Phi , \label{HAM}
\end{align}
where $\mu$ is the reduced mass of the nucleus-electron composite system. Next we introduce a further simplification in our system motivated by a mathematical as well a physical reason. The former motivation is simply the fact that in this way we obtain a solution of the energy eigenvalue problem  in terms of closed  known functions. The second motivation stems from the advantages of considering Rydberg atoms in studies of atom-surface interactions. In this case the size of the atom is very large in comparison with the  nucleus-surface distance, which means that it is adequate to consider an expansion in powers of $b/r$. This is why here we deal with the zeroth order of this expansion which amounts to set $b=0$, i.e. to locate the nucleus exactly at the $\theta$-interface. A detailed study of these zeroth-order Rydberg states is beyond the scope of the present work. Some preliminary results in a realistic setting are reported in Refs. \cite{EPL,PRA}. According to the previous simplification, the electromagnetic potentials take the form
\begin{align}
\Phi (\textbf{r}) = \frac{4}{4+\tilde{\theta}^2}\frac{e}{r} \quad , \quad \textbf{A} (\textbf{r}) = \frac{2 e \tilde{\theta}}{4 +
 \tilde{\theta} ^{2}} \frac{y \hat{\textbf{e}} _{x} - x \hat{\textbf{e}} _{y}}{\rho ^{2}} \left( 1 - \frac{\vert z \vert}{r} \right) . \label{HpotAz,b=0}
\end{align}
The scalar potential corresponds to the one generated by a pointlike nucleus at the origin with effective charge $4e / (4 + \tilde{\theta} ^{2})$. The axial symmetry around the $z$-axis motivates the introduction of spherical coordinates which yields the even simpler form of the vector potential
\begin{align}
A _{\varphi} (\textbf{r}) = - \frac{2\tilde{\theta} e}{4+\tilde{\theta}^2} \frac{1}{r} \, \frac{1-\left|\cos \vartheta\right|}{\sin\vartheta}.
\label{POTFIN}
\end{align}

{\subsection{Comparison with the case when the electron moves in the field of a magnetic monopole}
\label{COMP_MON}

The physical situation described by the Hamiltonian (\ref{HAM}) corresponds to the motion of an electron in the presence of  a material  medium with constitutive relations (\ref{ec3}), satisfying $\nabla\cdot {\mathbf B}\equiv 0$ everywhere. This setting is  certainly completely different from that of an electron moving in the field  of a dyon (a charged  magnetic monopole source) which has been previously  studied in many references \cite{MONOPOLE,VILLALBA}. Nevertheless, the similarity of the vector potential (\ref{POTFIN}) with that of the monopole
\begin{align}
A^M_{\varphi} (\textbf{r}) =  g\frac{1}{r} \,\frac{1-\cos\vartheta}{\sin{\vartheta}}
\label{POTMON}
\end{align}
may lead to some confusion and  motivates us to provide a  
better  account of the differences involved. 

The appearance of $|\cos \vartheta|$ in Eq. (\ref{POTFIN}) as opposed to $\cos \vartheta$ in Eq. (\ref{POTMON}) is crucial.  This guarantees that $A_{\varphi} (\textbf{r})$ is finite in the whole interval $0\leq \vartheta \leq \pi$, yielding a vector potential which is nonsingular in 
$\vartheta$ together with a  magnetic field 
$\mathbf B$, satisfying $\nabla\cdot {\mathbf B}=0$ everywhere, which is
\begin{equation}
{\mathbf B}=\frac{g}{r^2}\left( H\left({\pi}/{2} -\vartheta \right)- H\left(\vartheta- {\pi}/{2} \right) \right){\hat {\mathbf r}}.
\end{equation}
The integration of ${\mathbf B}$ over a sphere of radius $ \epsilon \rightarrow 0$ centered at  the origin produces
\begin{equation}
\oint_{\epsilon \rightarrow 0} {\mathbf B}\cdot d{\mathbf S}=2\pi g \left( \int_0^{\pi/2} \sin\vartheta \, d\vartheta -\int_{\pi/2}^{\pi} \sin\vartheta \, d\vartheta \right)=0,
\end{equation}
showing that $\nabla \cdot {\mathbf B}=0$ at the origin.

On the contrary, the monopole vector potential $A^M_{\varphi} (\textbf{r})$ is singular at $\vartheta=\pi$, thus requiring the introduction of the Dirac string and yielding a singular magnetic field ${\mathbf B}^M$ such that $\nabla \cdot {\mathbf B}^M=4\pi g \delta^3({\mathbf r})$.
 
An additional difference between the two cases arises because $ {\partial A_{\varphi}({\mathbf r})}/{\partial \vartheta}$ is  discontinuous at $\vartheta= \pi/2$. This prevents us to introduce the modified angular momentum operator 
\begin{equation}
{{\mathbf {\bar L}}} ={{\mathbf {L}}}-e{\mathbf r}\times \mathbf{A}-eg{\mathbf{\hat r}}, \label{MODANGMOM}
\end{equation}
that provides an alternative way of dealing with the quantization of the angular sector of the theory. In the case where ${\mathbf A}$ is regular, continuous and with continuous first derivatives the operators ${\bar L}_i$ satisfy the standard angular momentum commutation relations  and also commute with the corresponding Hamiltonian. When dealing with a magnetic monopole it is possible to perform  a  regularization of  the otherwise singular operators $L_i$ and $H$, arising from  the singular potential vector $A^M_{\varphi}$, which preserve such commutation  properties among the 
${\bar L}_i$'s together with $H$ \cite{MONOPOLE}. 

 The angular momentum commutation relations among the operators ${\bar L}_i$ rest upon the equality
\begin{equation}
[L_i, U_j]-[L_j, U_i]= i\hbar \epsilon_{ijk}\left(U_k-\frac{x_k}{r}\right), \qquad    {\mathbf U}=\frac{1}{g}{\mathbf r}\times {\mathbf A}.
\label{COND_GEN}
\end{equation} 
Choosing $i=x, j=y$ together with the vector potential  (\ref{POTFIN}) we obtain
\begin{equation}
[L_x, U_y]-[L_y, U_x]= i\hbar \frac{\sin^2\vartheta-|\cos\vartheta|-\cos^2 \vartheta}{1+|\cos\vartheta|}, \label{COND1}
\end{equation}
\begin{equation}
i\hbar \left(U_z-\frac{z}{r}\right)=i\hbar\frac{\sin^2\vartheta-\cos \vartheta-|\cos\vartheta|cos\vartheta}{1+|\cos \vartheta|},
\label{COND2}
\end{equation}
which satisfy the condition (\ref{COND_GEN}) only for $0<\vartheta <\pi/2$ when $\cos\vartheta >0$. In the case of the monopole $|\cos\vartheta|$ is replaced by $\cos\vartheta$ and  (\ref{COND1}) equals (\ref{COND2}) for all $\vartheta$. The impossibility of introducing ${\bar L}_i$ in our case 
requires an alternative procedure for obtaining the quantization of the angular sector of the Schr\"{o}dinger equation, which will be presented  in Section \ref{ANG_EQ}.

\subsection{ The resulting equations}
Introducing the EM potentials $(\Phi, A _{\varphi} )$ in the Hamiltonian (\ref{HAM}) we obtain the time-independent Schr\"{o}dinger equation
\begin{align}
& \left\lbrace -\frac{1}{2\mu}\frac{1}{r^2}\frac{\partial}{\partial r}\left(r^2 \frac{\partial}{\partial r} \right) -\frac{1}{2\mu} \frac{1}{r^2\sin \vartheta}\frac{\partial}{\partial \vartheta}\left(\sin \vartheta \frac{\partial}{\partial \vartheta} \right) -\frac{1}{2\mu} \frac{1}{r^2 \sin^2 \vartheta}\frac{\partial^2}{\partial \varphi^2}\quad + \right. \notag \\ & \hspace{3cm} \left. \frac{2a}{\mu r^2} \left( \frac{1}{1 + \vert \cos \vartheta \vert}\right) \left(i \frac{\partial}{\partial \varphi} +2a\right) - \frac{2a^2}{\mu r^2}
-\frac{\tilde{e}^2}{r} \right\rbrace \psi\left(r,\vartheta,\varphi \right) =E\psi\left(r,\vartheta,\varphi \right), 
\label{schrodinger1}
\end{align}
where we have introduced the notation 
\begin{align}
\tilde{e} ^{2} = \frac{4}{4 + \tilde{\theta} ^{2}}e ^{2} \quad , \quad a = \frac{\tilde{\theta}}{4+\tilde{\theta} ^{2}} e ^{2} . \label{e-a}
\end{align}
In Fig. \ref{GraficasAyE} we show the behavior of $a$ and $\tilde{e}^2$ as functions of $\tilde{\theta}$, which are bounded. Taking the physical values for the parameters, the maximum value of  $a$ is $a _{\mbox{\scriptsize m}} = e ^{2} / 4 = 1.824 \times 10 ^{-3}$ for $\tilde{\theta} = 2$, while the maximum value for $\tilde{e} ^{2}$ is $e ^{2}$ when $\tilde{\theta} = 0$. In order to make clear and amplify the new features of our Schr\"{o}dinger equation (\ref{schrodinger1}) we will consider $a$ and $e^2$ as free parameters, including the cases $a = 1$ and $a = a _{\mbox{\scriptsize m}}$, together with the more realistic case $\tilde{\theta} = \alpha = e ^{2}$ for which $a = 1.33 \times 10 ^{-5}$.
\begin{figure}
\begin{center}
\includegraphics[width=0.48\linewidth]{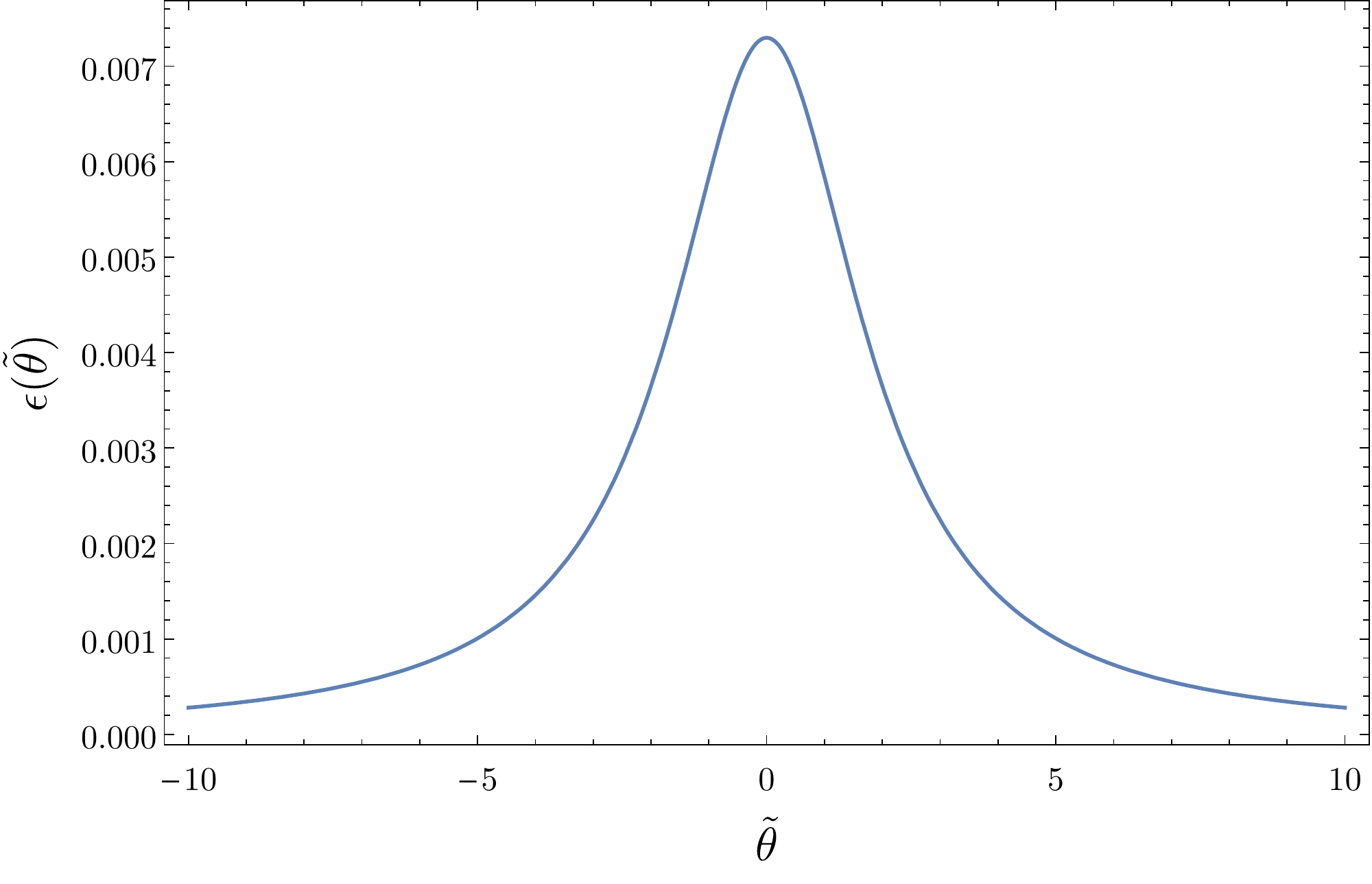}
\includegraphics[width=0.48\linewidth]{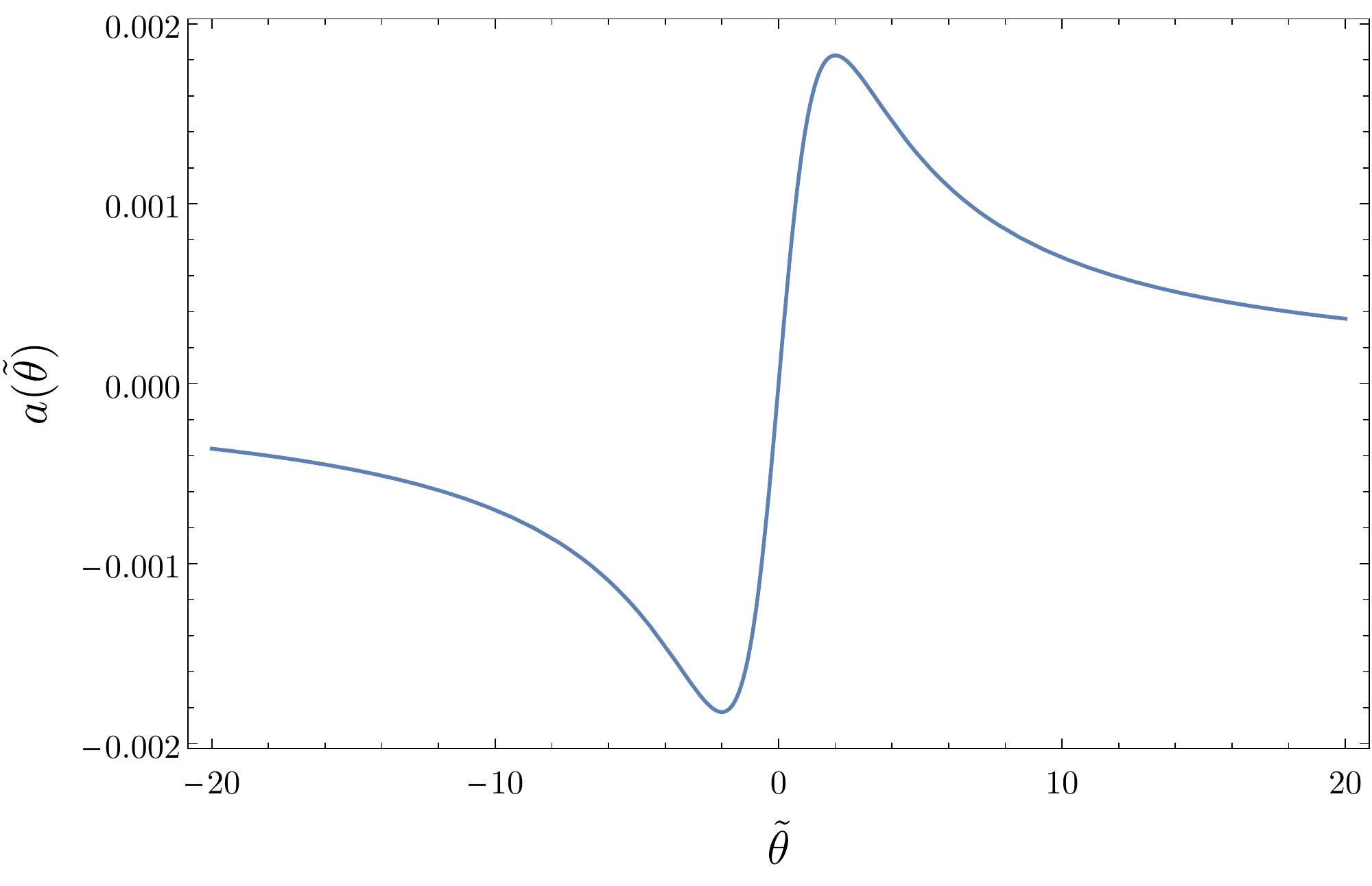}  
\end{center}
\caption{{ \protect \small 
Plots showing the parameters $\tilde{e}$ and $a$ as functions of   the dimensionless magnetoelectric polarization $\tilde{\theta}$,  according to Eq. (\ref{e-a}),  using $e^2=1/137$. The minimum (maximum) for $a$ occurs at  $\tilde{\theta}=-2$ ($\tilde{\theta}=2$).}}
\label{GraficasAyE}
\end{figure}

Now we concentrate on analyzing the solutions of the Schr\"{o}dinger equation (\ref{schrodinger1}) that describes the quantum-mechanical motion of the atomic electron. In the spherical coordinate system we can separate  variables and the wave function can be written as
\begin{align}
\psi\left(r,\vartheta,\varphi \right)  = \mathcal{R}(r)\Theta (\vartheta) F(\varphi). \label{soln1}
\end{align}
The axial symmetry of the problem leads to the conservation of $L _{z} = - i \partial / \partial \varphi$, in such a way that we can take $F(\varphi)\sim e^{im\varphi}$, with $m = 0, \pm 1, \pm 2,...$. The remaining functions $\mathcal{R}$ and $\Theta$ satisfy the differential equations
\begin{align}
\frac{1}{\Theta} \frac{1}{\sin \vartheta}\frac{d}{d \vartheta}\left(\sin \vartheta \frac{d \Theta}{d \vartheta} \right) - \frac{m^2}{\sin^2 \vartheta}- 4a \left( \frac{1}{1+ |\cos \vartheta|}\right) \left(2a-m\right) =-\lambda, \label{ec.angular1}
\end{align}
and
\begin{align}
\frac{1}{\mathcal{R}} \frac{d}{d r}\left(r ^{2} \frac{d \mathcal{R}}{d r} \right) + 4 a ^{2} + 2 \mu r \tilde{e} ^{2} + 2 \mu r ^{2} E = \lambda ,  \label{ec.radial1} 
\end{align}
where $\lambda$ is a constant of separation. Introducing the following definition 
\begin{align}
\mathcal{L} ^{2} = \textbf{L} ^{2} + 4a \left( \frac{2a - L _{z}}{1 + \vert \cos \vartheta \vert}\right), \label{L}
\end{align}
the angular equation  (\ref{ec.angular1}) reads $\mathcal{L} ^2 \Theta = \lambda \Theta$, which identifies $\lambda$ as the eigenvalues of the extended squared angular momentum. As usual we denote by $l(l+1)$ the eigenvalues of $\textbf{L} ^{2}$, with $l=0,1,2,\dots$, and we find it convenient to introduce the notation $\lambda = L (L+1)$. When $a=0$ we obtain $L=l$, so that $L$ can be considered as a deformation of the angular momentum quantum number $l$. In the next sections we solve the resulting differential equations (\ref{ec.angular1}) and (\ref{ec.radial1}).

\section{Solution of the angular equation}
\label{ANG_EQ}

We proceed by first computing the eigenfunctions of the generalized squared angular momentum operator $\mathcal{L}^2$. Introducing $x = \cos\vartheta$ together with $\Theta (\vartheta) = P (x)$, the differential equation (\ref{ec.angular1}) becomes
\begin{align}
\Bigg[ \left(1 - x ^{2} \right) \frac{d ^{2}}{d x ^{2}} - 2 x \frac{d}{d x} + \left( \lambda -\frac{m ^{2}}{1 - x ^{2}} - \frac{8 a ^{2} - 4 m a}{1 + \vert x \vert} \right) \Bigg] P (x) = 0 , \label{ec.angular2}
\end{align}
which holds for the interval $[-1,1]$. We observe that this equation depends only on the parameter $a$, which produces a deformation of the associated Legendre differential equation for $\lambda=n(n+1)$, which is recovered when $a=0$. We will call the solutions $P (x)$ of Eq. (\ref{ec.angular2}) the deformed associated Legendre functions.

Since the operator acting on $P$ is even under the parity operation  $x \rightarrow -x$, we can choose the functions $P (x)$ to be even or odd under this transformation. Let us denote by $P(x)$ ($P(-x)$) the solutions of Eq. (\ref{ec.angular2}) in the intervals $[0,1]$ ($[-1,0]$), respectively.  Notice that $P(x)$ has no definite parity and can be viewed as a general series expansion in powers of $x$. We construct the even and odd solutions of Eq. (\ref{ec.angular2}) by taking
\begin{align}
\mathcal{P} _{\rm even}(x) = H(x) P(x) + H(-x) P(-x) ,  \label{Solpar} \\
 \mathcal{P} _{\rm odd}(x) = H(x) P(x) - H(-x) P(-x) , \label{Solimpar}
\end{align}
and demanding both solutions, together with  their derivatives, to be continuous at $x=0$. These requirements impose the further conditions \cite{Schwinger} 
\begin{align}
\mathcal{P} _{\rm odd}(x \rightarrow 0_{\pm}) = 0 \quad , \quad \frac{d \mathcal{P} _{\rm even}}{d x} \Big| _{x\rightarrow 0_{\pm}} = 0 . \label{condparidad}
\end{align}
Next we study the solution $P(x)$ around the regular singular points $x=\pm 1$ and find that their behavior can be isolated in the following way 
\begin{align}
P(x) = (1-x) ^{\frac{\vert m \vert}{2}}(1+x)^{+\frac{1}{2} \vert m-4a \vert} f(x),
\label{FACTORP}
\end{align} 
where the differential equation defining $f(x)$ is to be  subsequently determined. 

\subsection{The solutions $P(x)$ in the interval $[0,1]$}

\label{ANG_EQ1}

In the above equation (\ref{FACTORP}) we have to  distinguish three cases: (1) $m>4a>0$, (2) $4a>m>0$ and (3) $0 > m$. We label the corresponding solutions   by $P _{1} (x)$, $P _{2} (x)$ and $P _{3} (x)$ respectively.

Let us first consider the case (1) $m>4a>0$, for which the function $P _{1}(x)$ reads
\begin{equation}
P _{1} (x) = (1-x) ^{\frac{m}{2}}(1+x)^{\frac{1}{2}{\left(m-4a\right)}} f _{1} (x) . \label{SolnP,m>0a}
\end{equation} 
Substituting this into (\ref{ec.angular2}) and introducing the new variable $u = (1-x)/2$, we get
\begin{equation}
u(1-u) \frac{d ^{2} f _{1}}{d u ^{2}} + \left\lbrace 1 + m - (2-4a+2m) u \right\rbrace\frac{d f _{1}}{d u} - \left\lbrace (m-2a) ^{2} + m - 2a - \lambda \right\rbrace f _{1} = 0 , \label{d2ga}
\end{equation}
which can be recognized as the standard Gauss hypergeometric equation
\begin{equation}
u(1-u) \frac{d ^{2} F}{d u ^{2}} +\left\lbrace C-(A+B+1) u \right\rbrace\frac{d F}{d u} - A B \; F=0, \label{hipergeoma}
\end{equation}  
with the solution ${}_{2} F _{1} (A,B;C;u)$ in standard notation \cite{Arfken,Abramowitz}. A direct comparison between Eqs. (\ref{hipergeoma}) and (\ref{d2ga}) reveals that $C = m+1$, $A+B+1 = 2-4a+2m$ and $AB = (m-2a)^2+m-2a- \lambda$. It is well known that the hypergeometric function is finite if it is a polynomial with a finite number of terms, and this is satisfied if $A$ or $B$ is a negative integer. If we fix $A = - n$, with $n = 0,1,2,...$, then we find that the constant of separation becomes
\begin{align}
\lambda = n (n+1) + (m - 2a) ^{2} + 2n(m - 2a) + (m - 2a) ,
\end{align}
which can be written in the form $L(L+1)$, with $L = n + m - 2a$. This analysis justifies our choice $\lambda = L(L+1)$. Identifying the parameters $A,B$ and $C$ from Eqs. (\ref{d2ga}) and (\ref{hipergeoma}) we find
\begin{equation}
A=\frac{1}{2}-2a+m+\frac{1}{2}\sqrt{1+4\lambda},\qquad 
B=\frac{1}{2}-2a+m-\frac{1}{2}\sqrt{1+4\lambda}, \qquad 
C=1+m , \label{ABa}
\end{equation}
and then the required solution is
\begin{equation}
f_1(x) = {}_2F_1\left(1-2a+m+L,m-2a-L;1+m;\frac{1-x}{2} \right),
\end{equation}
where we have substituted the parametrization $\lambda = L(L+1)$. Thus, up to a normalization factor, the full solution of Eq. (\ref{ec.angular2}) in the case (1) $m>4a>0$ is
\begin{equation}
P_1(x)=K (-1)^m(1-x)^{\frac{m}{2}}(1+x)^{\frac{1}{2}(m-4a)}{}_2F_1\left(1-2a+m+L,m-2a-L;1+m;\frac{1-x}{2} \right),  \label{SolPx,m>0}
\end{equation}
where $K$ is the aforementioned normalization constant and the factor  $(-1)^m $ has  been added to adopt the phase convention of Condon and Shortley \cite{Arfken}.

Following similar steps, one can further see that the solution for the case (2) $4a>m>0$ is 
\begin{equation}
P_2(x)=K (-1)^m(1-x)^{\frac{m}{2}}(1+x)^{-\frac{1}{2}(m-4a)}{}_2F_1\left(1+2a+L,2a-L;1+m;\frac{1-x}{2} \right),  
\label{SolPx,m<4a}
\end{equation}
and 
\begin{equation}
P_3(x)=K(1-x)^{-\frac{m}{2}}(1+x)^{-\frac{1}{2}(m-4a)} {}_2F_1\left(1+2a-m+L,2a-m-L;1-m;\frac{1-x}{2} \right),  \label{SolPx,m<0}
\end{equation}
for the last case (3) $0 \geq m$. At this stage, we only have to impose the continuity conditions (\ref{condparidad}) in $x=0$ for each case in order to have the full solutions according to Eqs. (\ref{Solpar}) and (\ref{Solimpar}).

\begin{table}[ht]
\begin{center}
\begin{tabular}{|c||c|c|c|c|c|c|c|c|c|c|}
        \hline
        $m$ &\multicolumn{10}{|c|}{$L$}\\
        \cline{2-11}
		\hline \hline
		0& 1.8696 & 2.1182  & 2.9051 & 3.6517 & 4.5417 & 5.4415 & 6.3831& 7.3307 & 8.2956 &  9.2637  \\
		\hline
		-1&--& 2.8207 & 3.1998 & 4.7776 & 5.6661& 6.5623 & 7.4973 & 8.4377 & 9.3959&   10.3575  \\
		\hline
		-2&--& -- & 3.8011 & 4.2603 &  5.8696 & 6.7596  & 7.6558 & 8.5876&  9.5244 & 10.4786  \\
		\hline
		-3&--& --& -- & 4.7935 & 5.3077 &  8.7313 & 9.6618 & 10.5967 & 11.5486 & 12.5035 \\
		\hline
		-4&--& -- & -- & -- & 5.7912 & 6.3463 & 9.7943 & 10.7244 & 11.6585 & 12.6089 \\
		\hline
		-5&--& -- & -- & -- & -- & 6.7916 & 7.3785 & 10.8480 & 11.7783 & 12.7122  \\
		\hline
		-6&--& -- & -- & -- & -- & -- & 7.7935 & 8.4060  & 11.8946 & 12.8254 \\
		\hline
		-7&--& -- & -- & -- & -- & -- & -- & 8.7960 & 9.4299 &   12.9356 \\
		\hline
		-8&--& -- & -- & -- & -- & -- & -- & -- & 9.7989 & 10.4510 \\
		\hline
		-9&--& -- & -- & -- & -- & -- & -- & -- & -- & 10.8020 \\
		\hline		
\hline
1& 1.7861 & 2.4567 & 3.3585 & 4.2731 & 5.2299 & 6.1925 & 7.1691 & 8.1483 & 9.1337 & 10.1205\\
\hline
2& -- & 2.0000 & 3.0000 & 4.0000 & 5.0000 & 6.0000 & 7.0000 & 8.0000 & 9.0000 & 10.0000  \\
\hline
3& --& -- & 2.5053 & 3.6865 & 4.7471 & 5.7950 & 6.8226 & 7.8463 & 8.8623 & 9.8767 \\
\hline
4& -- & -- & -- & 3.1990 & 4.4648 & 5.5599 & 6.6357 & 7.6811 & 8.7204 & 9.7477\\
\hline
5& -- & -- & -- & -- & 3.9999 & 5.3017 & 6.4160 & 7.5084 & 8.5654 & 9.6152 \\
\hline
6& -- & -- & -- & -- & -- & 4.8627 & 6.1772 & 7.3019 & 8.4043 & 9.4689\\
\hline
7& -- & -- & -- & -- & -- & -- & 5.7628 & 7.0792 & 8.2091 & 9.3173\\
\hline
8& -- & -- & -- & -- & -- & -- & -- & 6.6870 & 7.9999 & 9.1320\\
\hline
9& -- & -- & -- & -- & -- & -- & -- & -- & 7.6275  & 8.9344\\
\hline 
\end{tabular}
\caption{{\protect \small Numerical values for $L$  within the range  $-9<m<+9$,
for the choice $a=1$.}}
\label{tabla2a}
\end{center}
\end{table}
\begin{table}[ht]
\begin{center}
\begin{tabular}{|c||c|c|c|c|c|c|c|c|c|c|}
        \hline
        $m$ &\multicolumn{10}{|c|}{$L$}\\
        \cline{2-11}
	    
		\hline \hline
		0& 0.000018 & 1.000005  & 2.000003 & 3.000002 & 4.000002 & 5.000002 & 6.000001 & 7.000001 & 8.0000009 &  9.000001  \\
		\hline
		-1&--& 1.0018 & 2.0009 & 3.0006 & 4.0005 & 5.0004 & 6.0004 & 7.0003& 8.00002 &   9.0002  \\
		\hline
		-2&--& -- & 2.0022 & 3.0013 &  4.0010 & 5.0008  & 6.0007 & 7.0006 &  8.0005 & 9.0005 \\
		\hline
		-3&--& --& -- & 3.0025 & 4.0016 & 5.0013 & 6.0011 & 7.0009 & 8.0008 & 9.0007 \\
		\hline
		-4&--& -- & -- & -- & 4.0026 & 5.0018 & 6.0015 & 7.0012 & 8.0011 & 9.0010 \\
		\hline
		-5&--& -- & -- & -- & -- & 5.0027 & 6.0020 & 7.0017 & 8.0014 & 9.0012  \\
		\hline
		-6&--& -- & -- & -- & -- & -- & 6.0028 & 7.0021  & 8.0018 & 9.0016 \\
		\hline
		-7&--& -- & -- & -- & -- & -- & -- & 7.0028 & 8.0022 &   9.0019   \\
		\hline
		-8&--& -- & -- & -- & -- & -- & -- & -- & 8.0029 & 9.0023  \\
		\hline
		-9&--& -- & -- & -- & -- & -- & -- & -- & -- & 9.0029 \\
		\hline
		\hline 
		1& 0.99818 & 1.9990 & 2.9993 & 3.9994 & 4.9996 & 5.9996 & 6.9996 & 7.9997 & 8.9997 & 9.9997\\
		\hline
		2& -- & 1.9977 & 2.9986 & 3.9989 & 4.9991 & 5.9992 & 6.9993 & 7.9994 & 8.9995 & 9.9995  \\
		\hline
		3& --& -- & 2.9974 & 3.9983 & 4.9986 & 5.9988 & 6.9990 & 7.9991 & 8.9992 & 9.9993 \\
		\hline
		4& -- & -- & -- & 3.9973 & 4.9981 & 5.9984 & 6.9987 & 7.9988 & 8.9989 & 9.9990\\
		\hline
		5& -- & -- & -- & -- & 4.9972 & 5.9979 & 6.9982 & 7.9985 & 8.9987 & 9.9988 \\
		\hline
		6& -- & -- & -- & -- & -- & 5.9971 & 6.9978 & 7.9981 & 8.9984  & 9.9985\\
		\hline
		7& -- & -- & -- & -- & -- & -- & 6.9971 & 7.9977 & 8.9980 & 9.9983\\
		\hline
		8& -- & -- & -- & -- & -- & -- & -- & 7.9970 & 8.9977 & 9.9979\\
		\hline
		9& -- & -- & -- & -- & -- & -- & -- & -- & 8.9970  & 9.9976\\
		\hline
		10& -- & -- & -- & -- & -- & -- & -- & -- & -- &  9.9969 \\
		\hline
\end{tabular}
\caption{{\protect \small Numerical values for $L$  within the range  $-9<m<+9$,
for the choice $a=a_{\mbox{\scriptsize m}}=1.82\times10^{-3}$.}} 
\label{tabla1b}	
\end{center}
\end{table}
\begin{table}
\begin{center}
\begin{tabular}{|c||c|c|c|c|c|c|c|c|c|c|}
        \hline
         &\multicolumn{10}{|c|}{$l$}\\
        \cline{2-11}
		$m$ & 0& 1 & 2 & 3 & 4 & 5 & 6 & 7 & 8 & 9  \\ 
		\hline \hline
		0& 5.5452 $a^2$ & 1.5452 $a^2$  & 1.0452$a^2$& 0.7119$a^2$ & 0.5702$a^2$& 0.4578$a^2$ & 0.4048$a^2$& 0.3378$a^2$ & 0.3004$a^2$&  0.2672 $a^2$ \\
		\hline
			$\mp$ 1&--& $\pm$ 1.0000 & $\pm$ 0.5000 & $\pm$ 0.3750 & $\pm$ 0.2813 & $\pm$ 0.2344 & $\pm$ 0.1953 & $\pm$ 0.1709 & $\pm$ 0.1495 & $\pm$ 0.1346 \\
				\hline
				$\mp$ 2&--& -- & $\pm$1.2500 & $\pm$0.7500 & $\pm$0.5938 & $\pm$0.4688 & $\pm$0.4004 & $\pm$0.3418 & $\pm$0.3038& $\pm$0.2692  \\
				\hline
				$\mp$3&--& --& -- & $\pm$1.3750 & $\pm$0.9063 & $\pm$0.7422 & $\pm$0.6055 & $\pm$0.5264 & $\pm$0.4572 & $\pm$0.4102  \\
				\hline
				$\mp$4&--& -- & -- & -- & $\pm$1.4531 & $\pm$1.0156 & $\pm$0.8516 & $\pm$0.7109 & $\pm$0.6263 & $\pm$0.5511 \\
				\hline
				$\mp$5&--& -- & -- & -- & -- & $\pm$1.5078 & $\pm$1.0977 & $\pm$0.9365 & $\pm$0.7955 & $\pm$0.7083  \\
				\hline
				$\mp$6&--& -- & -- & -- & -- & -- & $\pm$1.5488 & $\pm$1.1621 & $\pm$1.0050 & $\pm$0.8654 \\
				\hline
				$\mp$7&--& -- & -- & -- & -- & -- & -- & $\pm$1.5811 & $\pm$1.2144 & $\pm$1.0617 \\
				\hline
				$\mp$8&--& -- & -- & -- & -- & -- & -- & -- & $\pm$1.6072 & $\pm$1.2581 \\
				\hline
				$\mp$9&--& -- & -- & -- & -- & -- & -- & -- & -- & $\pm$1.6291  \\
				\hline
\end{tabular}
\caption{{\protect \small Numerical values of the coefficient  $F(l,m)$ in Eq. (\ref{EXPANDL}), for a $\theta$-vacuum with  $a=1,3313\times 10^{-5}$. Notice that for  $m=0$  the differences $L-l$ are of order $a^2$. In the row $m=0$ we report $(L-l)$.}}
\label{tabla1}	
\end{center}
\end{table}

\subsection{The quantization of $L$}

In the problem at hand, the continuity conditions at $x=0$ required by the construction of angular wave functions with definite parity (\ref{condparidad}) produce the quantization of the quantum number $L$. Again, we have the three cases defined in  section \ref{ANG_EQ1}. Here we list the required conditions, which can only be solved numerically for each value of $m$ and a given choice of $a$.

\subsubsection*{Case (1) $m>4a>0$} 
The equation yielding the eigenvalues of $L$ for the odd solution arises from the condition $P_1(x \rightarrow 0_+)=0$, which  is
\begin{equation}
{}_2F_1\left(1-2a+m+L,m-2a-L;1+m;\frac{1}{2} \right)  =0, \label{Fimpares}
\end{equation}
while the even case requires $[dP_1/dx ](x \rightarrow 0_+)=0$, yielding
\begin{multline}
-2a\,\,{}_2F_1\left(m-2a+L+1,m-2a-L;m+1;\frac{1}{2} \right) -\frac{(m-2a+L+1)(m-2a-L)}{2(1+m)}\times \nonumber \\
 {}_2F_1\left(m-2a+L+2,m-2a-L+1;m+2;\frac{1}{2} \right)=0. \label{Fpares}
\end{multline}

\subsubsection*{Case (2) $4a>m>0$}

Similarly to the previous case, for the odd solution we must satisfy 
\begin{equation}
{}_2F_1\left(1+2a+L,2a-L;1+m;\frac{1}{2} \right)  =0. \label{Fimpares,m<4a}
\end{equation}
while the condition for the  even case is
\begin{eqnarray}
(2a-m)\,\,{}_2F_1\left(2a+L+1,2a-L;m+1;\frac{1}{2} \right) -\frac{(2 a - L) (1 + 2 a + L)}{2(1+m)}\times\\
	{}_2F_1\left(2a+L+2,2a-L+1;m+2;\frac{1}{2} \right)=0. \label{Fpares,m<4a}
\end{eqnarray}

\subsubsection*{Case (3) $m<0$}
Finally, for the odd solution now we have
\begin{equation}
{}_2F_1\left(1+2a-m+L,2a-m-L;1-m;\frac{1}{2} \right)  =0. \label{Fimparesm<0}
\end{equation}
while the even case requires
\begin{multline}
2a\,\,{}_2F_1\left(2a-m+L+1,2a-m-L;1-m;\frac{1}{2} \right)-\frac{(2a-m+L+1)(2a-m-L)}{2(1-m)}\times \\
	{}_2F_1\left(2a-m+L+2,2a-m-L+1;2-m;\frac{1}{2} \right)=0. \label{Fparesm<0}
\end{multline}

\begin{figure}
\begin{center}
\includegraphics[width=0.44\linewidth]{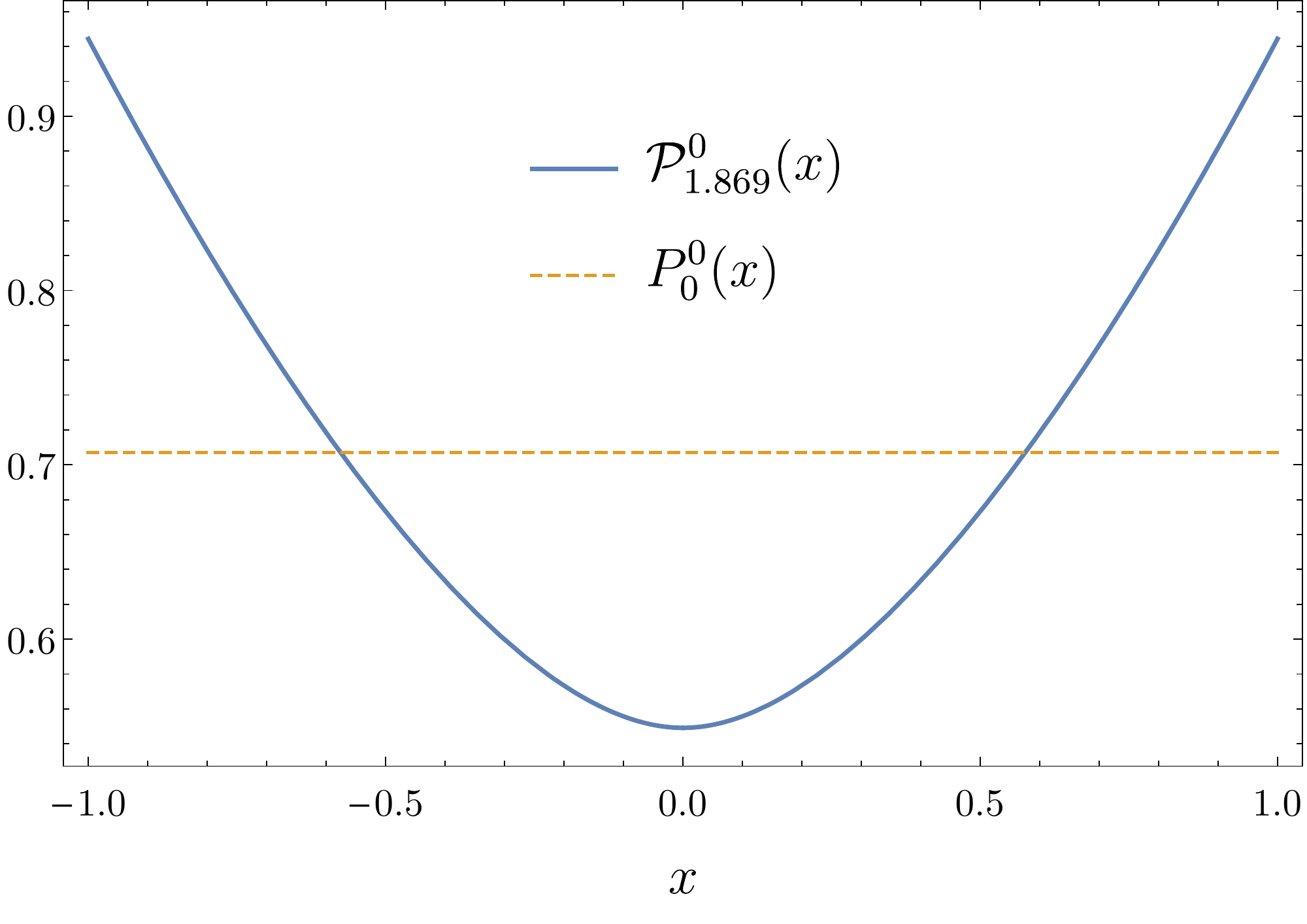}
\includegraphics[width=0.44\linewidth]{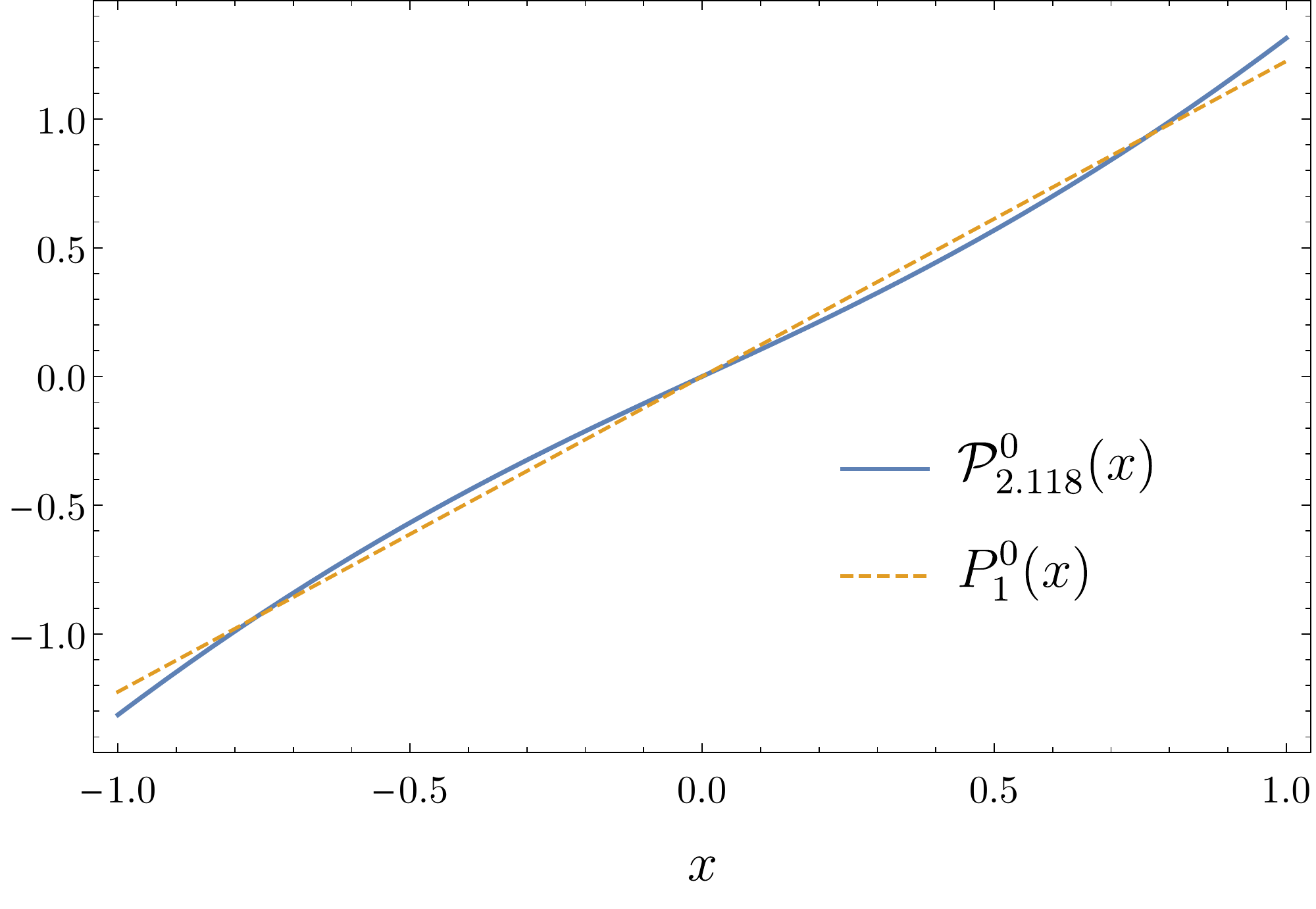}  
\includegraphics[width=0.44\linewidth]{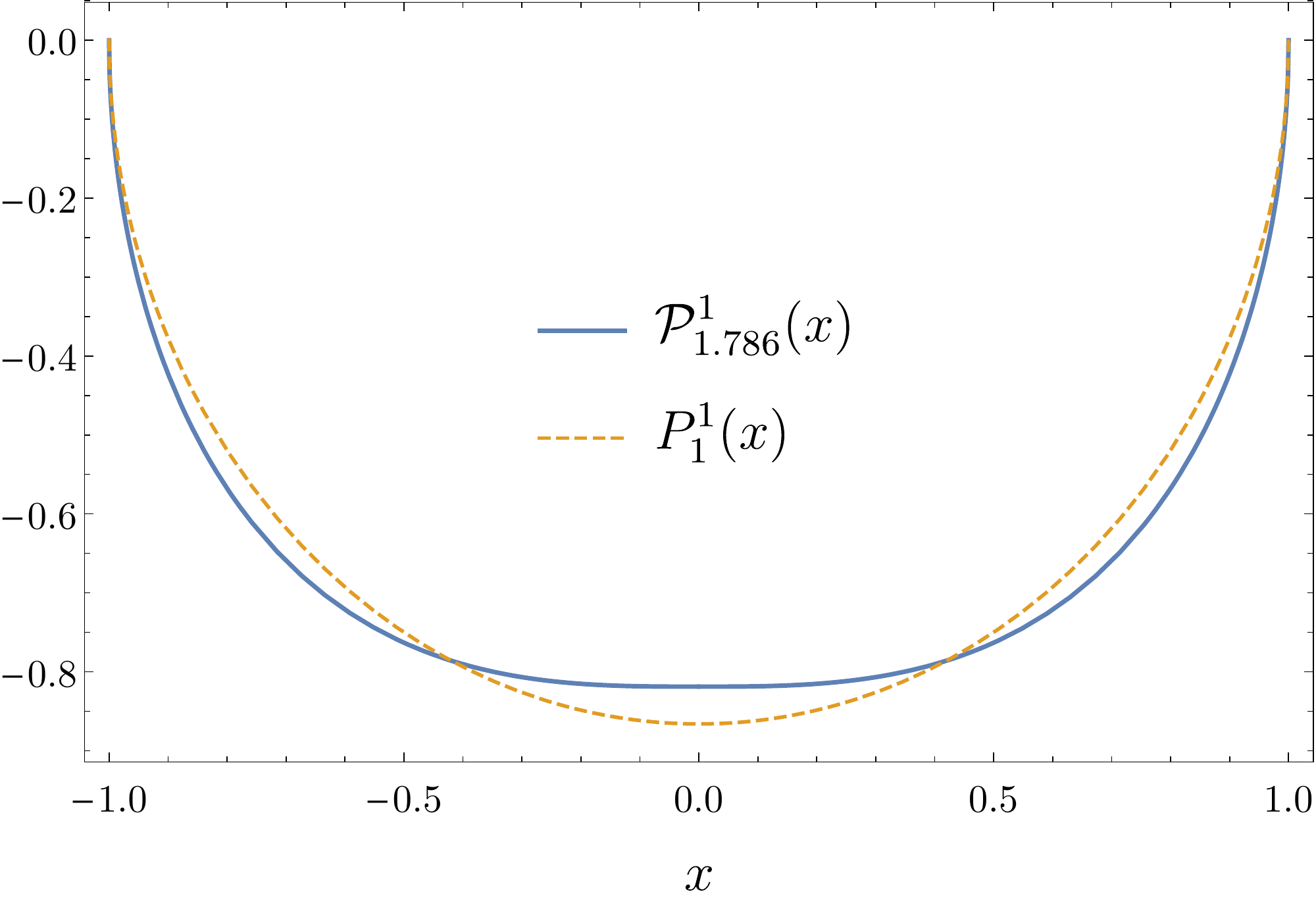}
\includegraphics[width=0.44\linewidth]{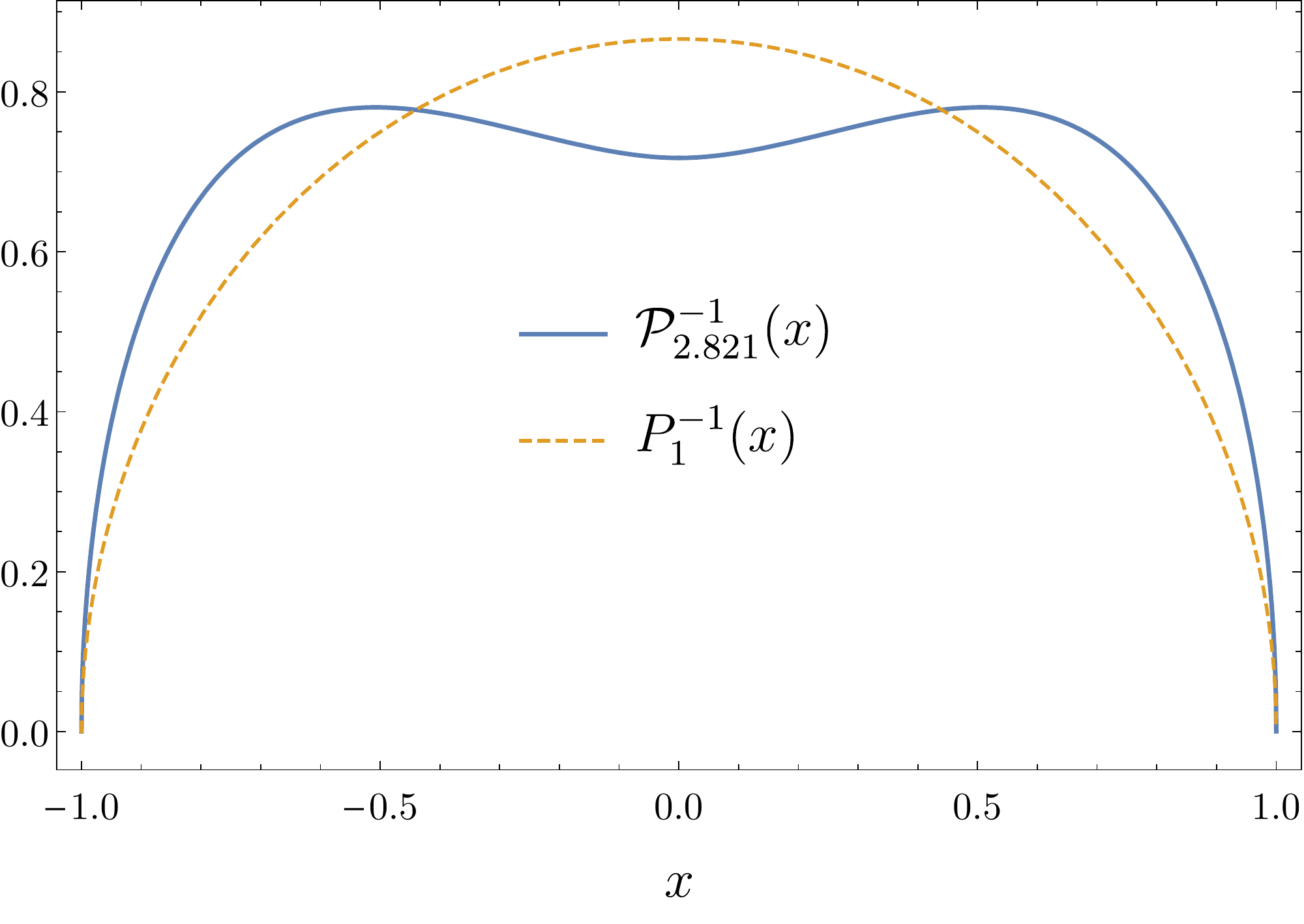}
\includegraphics[width=0.44\linewidth]{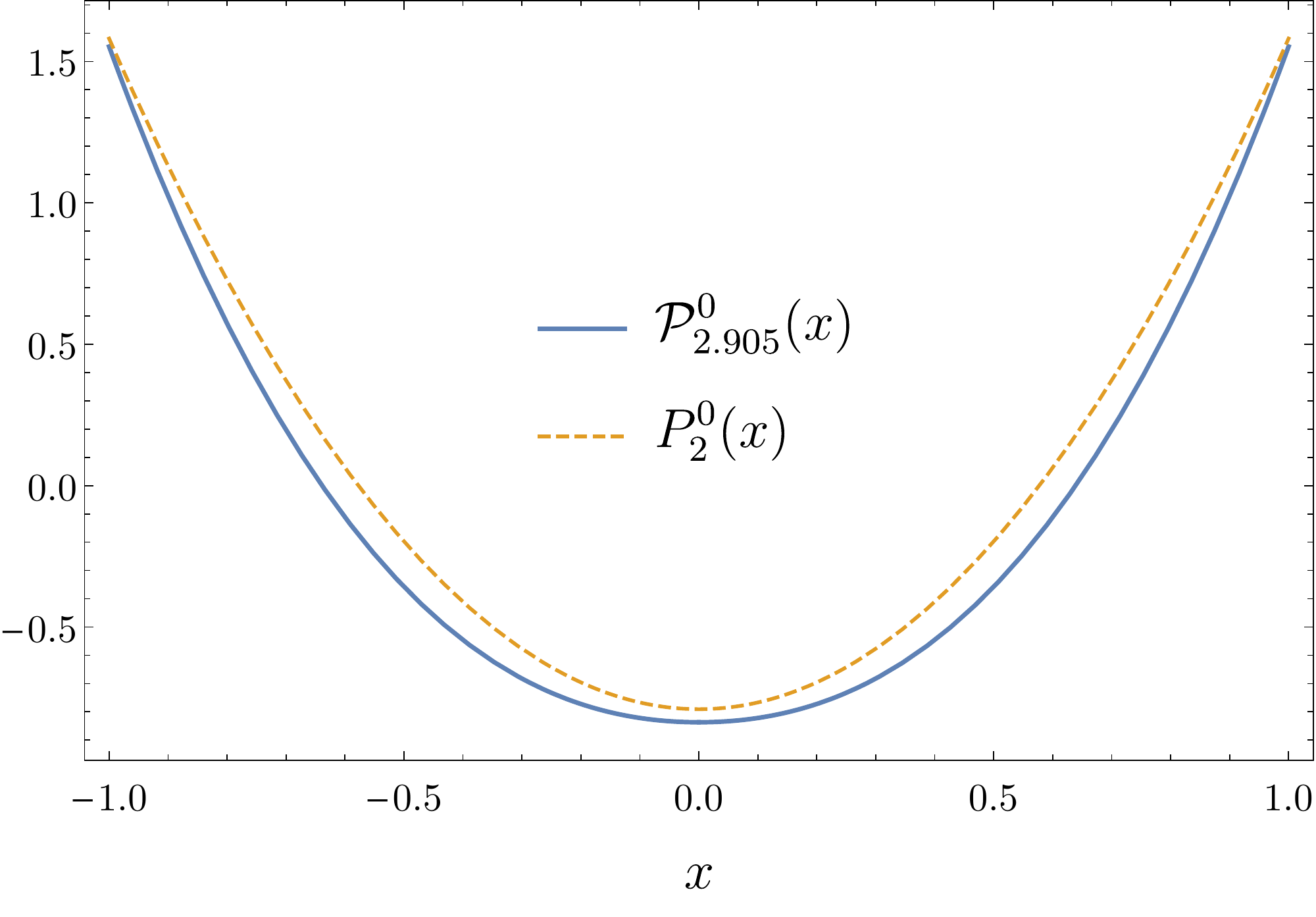} 
\includegraphics[width=0.44\linewidth]{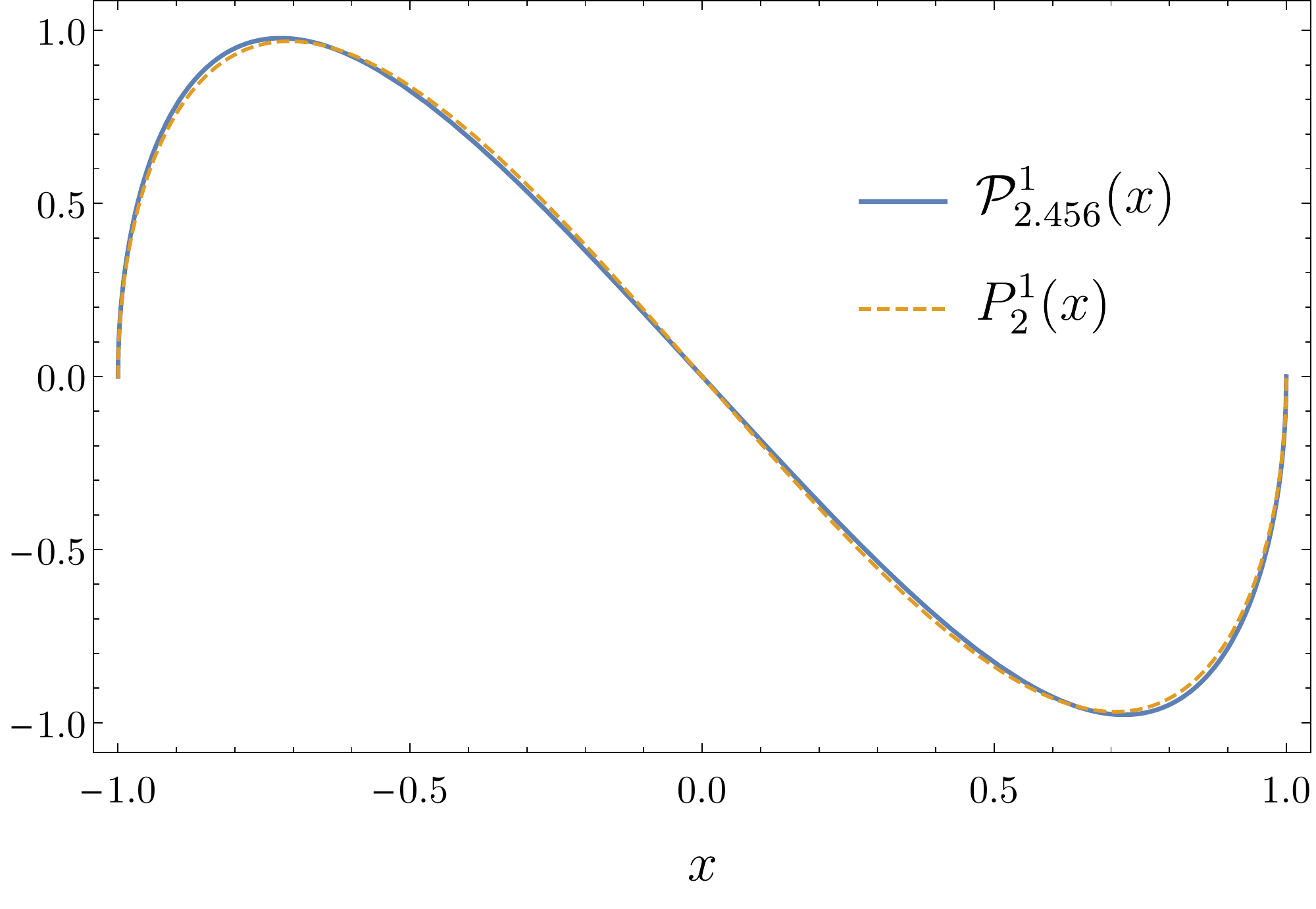} 
\includegraphics[width=0.44\linewidth]{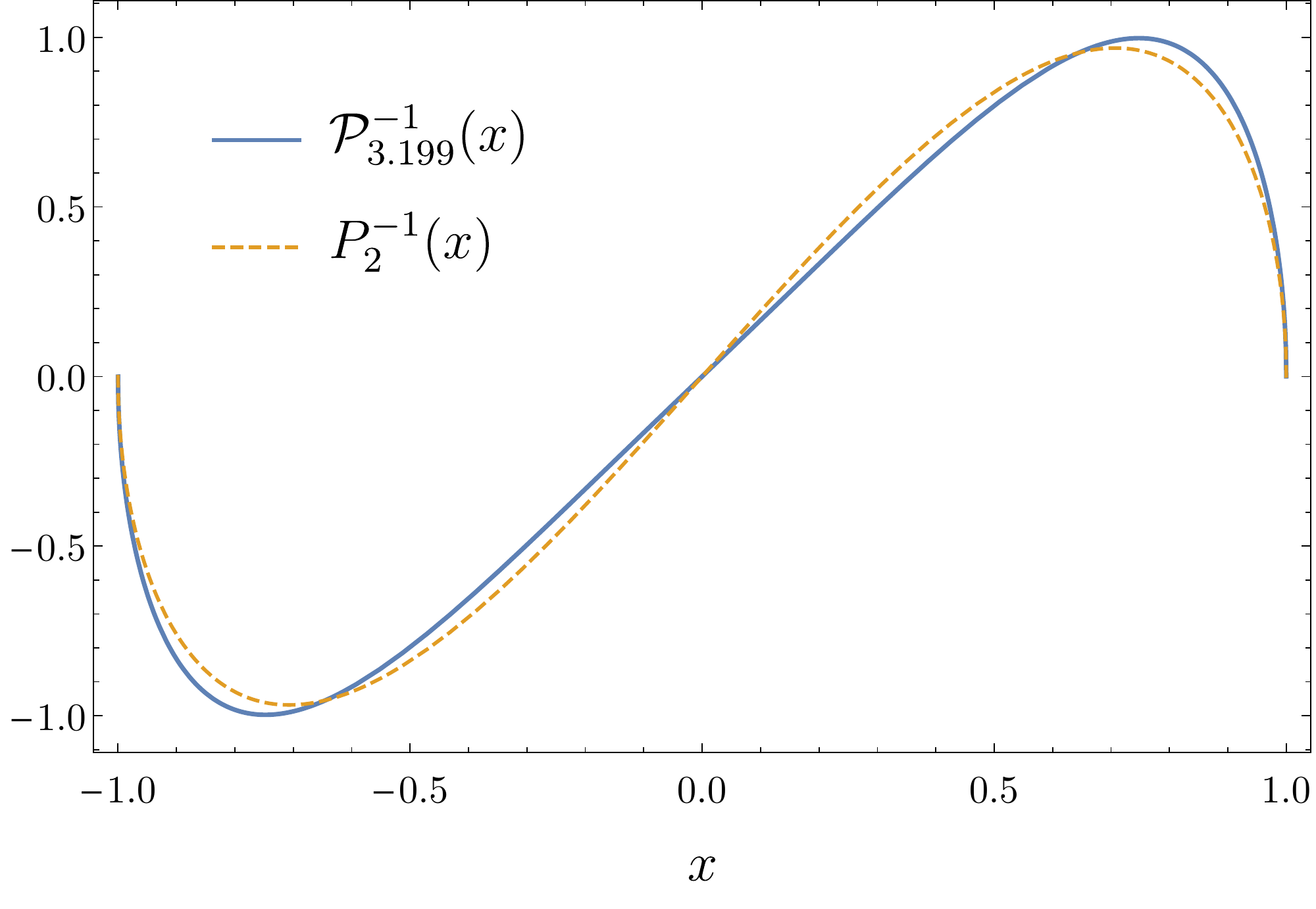} 
\includegraphics[width=0.44\linewidth]{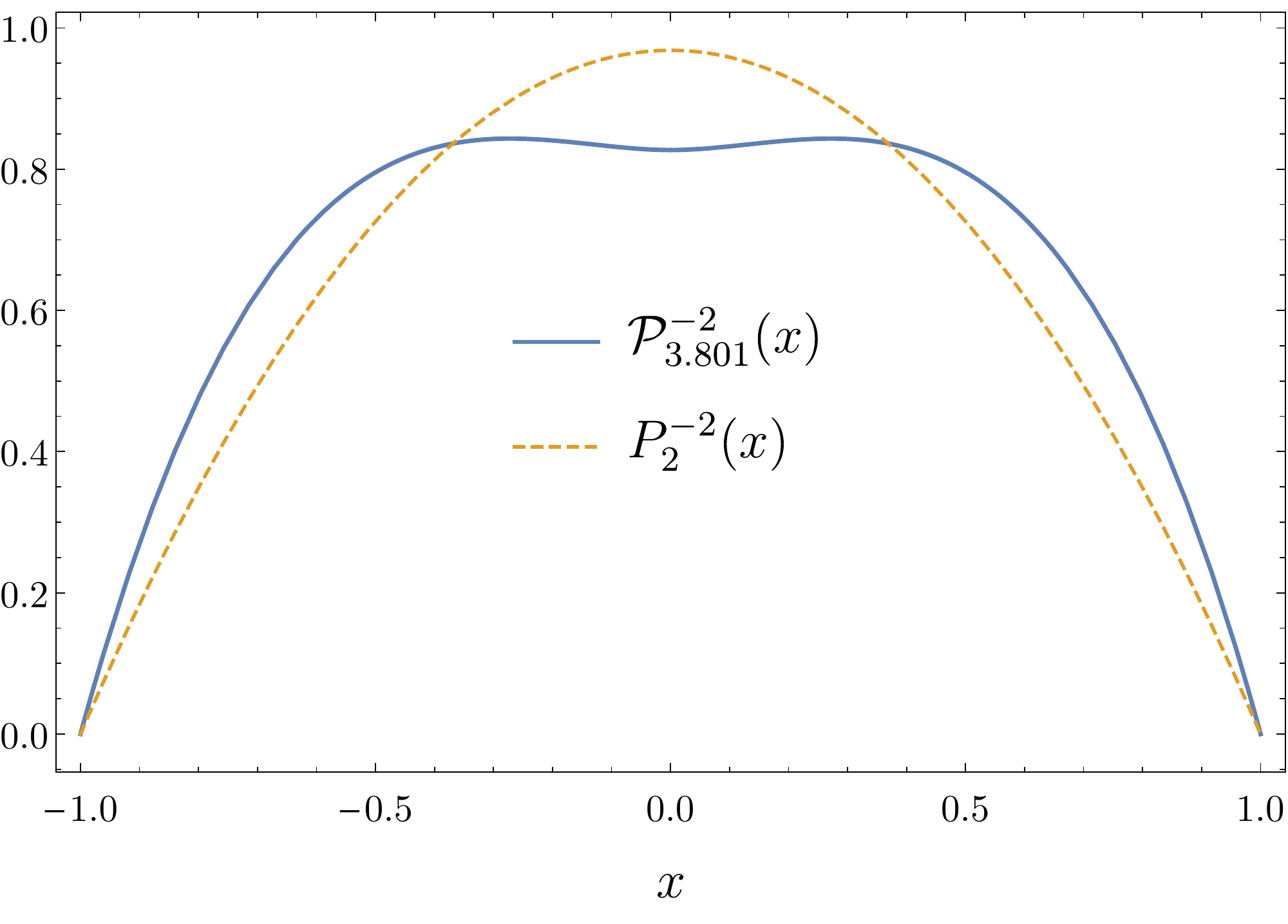}
 \end{center}
\caption{{\protect \small Plots of some lower order functions $\mathcal{P}_{L}^{m}(x)$ (solid line), for  $a=1$, together with the corresponding associated Legendre function $P^{m}_{l}(x)$ (dashed line).}}
\label{Figura}
\end{figure}
Table \ref{tabla2a} shows the numerical solutions for the deformed angular momentum $L$ in the case of $a = 1$, which is taken to exhibit drastic changes in comparison with the standard case $ a = 0$. The analogous results for the maximal case $a _{\mbox{\scriptsize m}} = 1.824 \times 10 ^{-3}$ are presented in Table \ref{tabla1b}, where we see that such value of $a$ it is already enough to wipe out the deformations in $L$, yielding $l$ as a very good approximation. This situation in even more severe in the case where $\tilde{\theta} =  \alpha$ and $a = \alpha ^{2} / (4 + \alpha ^{2} ) = 1.33 \times 10 ^{-5}$. In both cases we can expand $L$ to first order in $a$ as
\begin{align}
L = l + F(l,m) a \label{EXPANDL}
\end{align}
and present the results in terms of the function $F(l,m)$. In the case $m=0$ we find that the difference $(L-l)$ is of order $a ^{2}$. The numerical values are reported in Table \ref{tabla1} 

In this way, the deformed associated Legendre functions, which are the solutions of Eq. (\ref{ec.angular2}), are labeled by ${\mathcal P}_L^m$ and are constructed according to the definitions (\ref{Solpar}) and (\ref{Solimpar}). In Fig. \ref{Figura} we plot some examples of the functions ${\mathcal P}_L^m$, together with their corresponding undeformed associated Legendre functions $P_l^m(x)$ for the case $a=1$, where the differences become noticeable. The normalization of the functions  ${\mathcal P}_L^m$ has been computed numerically.

\section{Solution of the radial equation}

\label{RAD_EQ}

Going back to the radial differential equation (\ref{ec.radial1}), we define $\mathcal{R} \equiv  U(r)/r$ to obtain
\begin{align}
\frac{d ^{2}}{d r ^{2}} U(r)+ 2 \mu \left[ E +  \frac{\tilde{e} ^{2}}{r} - \frac{L(L+1) - 4 a ^{2}}{2 \mu r ^{2}} \right] U(r) = 0 , \label{ec.radial3}
\end{align}
where we have substituted $\lambda = L(L+1)$. In order to write Eq. (\ref{ec.radial3}) in an even more familiar form, it is convenient to define the function $\tilde \ell > 0$  as 
\begin{align}
\tilde{\ell}(\tilde{\ell} + 1 ) = L(L+1)- 4 a ^{2} ,\quad  \rightarrow \quad \tilde{\ell} = - \frac{1}{2} + \frac{1}{2} \sqrt{(2L+4a+1)(2L-4a+1)}, \label{ell}
\end{align} 
such that Eq. (\ref{ec.radial3}) takes the form of the radial equation for a hydrogen atom
\begin{align}
\frac{d ^{2}}{d r ^{2}} U(r) + 2 \mu \left[ E +  \frac{\tilde{e} ^{2}}{r} - \frac{\tilde{\ell}(\tilde{\ell} + 1)}{2 \mu r ^{2}} \right] U(r)=0. \label{ec.radial.ell}
\end{align}
In this way, the final solution of the radial differential  equation (\ref{ec.radial1}) is
\begin{align}
\mathcal{R}_{\nu,L, m}(r)= N\,\, \left(\frac{2r}{\tilde{a}_0 {\tilde n} } \right)^{\tilde{\ell}} \exp \left\lbrace -\frac{r}{\tilde{a}_0 {\tilde n}   }\right\rbrace \mathrm{L}_{\nu}^{2\tilde{\ell}+1}\left( \frac{2r}{\tilde{a}_0 {\tilde n} }\right), \label{Rnl}
\end{align}
Here  $L_\nu^\alpha$ are the associated  Laguerre polynomials of order ${\nu}$ and index $\alpha$, where $\nu=0,1,2, \dots$ indicates the number of nodes in the radial function,  ${\tilde n}=\nu+{\tilde \ell} +1$ is a deformation of the principal quantum number $n=\nu+l+1$ of the hydrogen atom and ${\tilde a}_0=1/(\mu {\tilde e}^2)$ is the modified Bohr radius. 
The energy eigenvalues are found to be
\begin{align}
E_{\nu,L, m}=-\frac{\mu\tilde{e}^4}{2(\nu+\tilde{\ell}+1)^2}, \label{Energia}
\end{align}
where  $\tilde{\ell}$ is as defined in Eq. (\ref{ell}). 

Next we calculate the lowest order correction in $\alpha=e^2$ to the energy shifts $E_{\nu,L,m}-E_{n,l,m}$, where $E_{n,l,m}=-\mu e^4/2n^2$. To this end it is convenient to  expand $E_{\nu,L,m}$ to order ${\tilde \theta}^2$. This is achieved by first expanding all quantities involved to order $a^2$. We add the next term corresponding to  Eq. (\ref{EXPANDL}) by writing  $L=l+F(l,m)a+ {\tilde G}(l,m) a^2 +\dots$. Let us observe that we have already identified numerically only  the coefficients $F(l,m)$ in Table \ref{tabla1}. Also, from Eq. (\ref{ell}) we obtain ${\tilde l}=L-4a^2/(2L+1)+\dots$ in such a way that we have
\begin{equation} 
{\tilde l}=l+F(l,m)a+G(l,m)a^2+\dots
\label{tildel}
\end{equation}
 to order $a^2$, where 
$G={\tilde G}-4/(2l+1)$. Using Eq. (\ref{tildel}) together with  the definitions of $\tilde{e}^2$ and  $a$ in Eqs. (\ref{e-a}) as functions of $\tilde \theta$, we expand Eq.  (\ref{Energia}) yielding
\begin{equation}
E_{\nu ,L,m}=-\frac{\mu e^{4}}{2n^{2}}+\frac{\mu e^{6}}{4n^{3}}%
F(l,m)\;\tilde{\theta}+\left( \frac{\mu e^{4}}{4n^{2}}-3\frac{\mu e^{8}}{32n^{4}}%
F^2(l,m)-\frac{1}{2l+1}\frac{\mu e^{8}}{4n^{3}}+\frac{\mu e^{8}}{16n^{3}}%
G(l,m)\right) \tilde{\theta}^{2}+ \dots, 
\label{EnergiaOrdenCuad}
\end{equation}
which shows the removal of the degeneracy of the original levels in  the hydrogen atom. In the case when $\tilde \theta = e ^{2} = \alpha$ the above expression simplifies, giving  
\begin{align}
 \frac{E _{n,l,m} - E _{n}}{E _{n}} = - \frac{1}{2} \left( 1 + \frac{F(l,m)}{n}\right) \alpha ^{2} \equiv \Delta \, {\tilde \theta} ^{2} .
\label{DELTAE}
\end{align}
to first order in $\alpha^2$.
The numerical results for the first energy levels are shown in Table \ref{tabla4}. Let us observe  that, to the order considered, for each value of $n$ the lines with $l, m=0$  remain degenerate. This is because $F(l,m=0)=0$, according to Table \ref{tabla1}, so that the corrections to the energies of the levels with  $m=0$ involve only the coefficient $G(l,m=0)$, which produces  a contribution of order $\alpha^4$ to the ratio (\ref{DELTAE}).  Within the approximation, we then take 
$\Delta=-0.5$  for these cases with $m=0$ in Eq. (\ref{DELTAE}). Nevertheless, we expect that  a further expansion in powers of $\alpha$ will completely remove such degeneracy. Accordingly, Figure \ref{NivelesEnergia} illustrates the energy splitting of the first three levels in the hydrogen atom.

We conclude this section with Fig. \ref{GraficaRadialA}, where we present plots of the probability density $\rho _{\nu,L,m}= 
\vert r \mathcal{R} _{\nu,L,m}(r) \vert ^{2}$ for the ground state and the first two excited states and compare them with the analogous unperturbed probability density $\vert r R _{n,l} \vert ^{2}$ of the hydrogen atom, which is independent of $m$. Each curve corresponds to an energy state characterized by a different value of $(\nu+L+1)$, where the dependence in $m$, which yields the corresponding $L$ in each wave function, is made explicit. Let us recall that we have $n=\nu+l+1$ for the hydrogen atom.   In order to amplify the effects of the $\theta$-vacuum we take $a=0.2$ and present the corresponding values of $L$ in Table \ref{tablaL}.

\begin{table}
\begin{center}
	\begin{tabular}{||c|c|c|c|c|c||}
		\hline\hline
		$n$ & $l$ & $m$ & $\Delta $ & $E_n$ (eV)& $E_{n,l,m}-E_n$ (meV)\\ 
		\hline \hline
	     1& 0 & 0 & -0.5000 & -13,6 &0,3621 \\ 
		\hline 
		\multirow{4}{1cm}{\centering 2} & 0 & 0 & -0.5000 & -3,4 & 0,0905\\
		\cline{2-6}
		& \multirow{3}{1cm}{\centering 1} & -1  & -0.7500 & -3,4 & 0,1358 \\ 
		\cline{3-6}
	    &  & 0   & -0.5000& -3,4 & 0,0905 \\
	    \cline{3-6}
	    &  & 1   & -0.2500& -3,4 & 0,0453\\
	    \hline 
	    \multirow{9}{1cm}{\centering 3} & 0 & 0   &   -0.5000 & -1.51 & 0,0402 \\
	    \cline{2-6} 
	    & \multirow{3}{1cm}{\centering 1} & -1  & -0.6667& -1.51 & 0,0536\\ 
	   	\cline{3-6}
	   	&  & 0   & -0.5000& -1.51 & 0,0402\\
	    \cline{3-6}
	   	&  & 1   & -0.3333& -1.51 & 0,0268\\
        \cline{2-6}
	   	& \multirow{5}{1cm}{\centering 2} & -2   & -0.7083& -1.51 & 0,0570 \\
	   	\cline{3-6}
	   	&  & -1   & -0.5833& -1.51 & 0,0469 \\
	   	\cline{3-6}
	   	&  & 0   & -0.5000& -1.51 & 0,0402\\
	   	\cline{3-6}
	   	&  & 1  & -0.4167& -1.51 & 0,0335\\
	   	\cline{3-6}
	   	&  & 2   & -0.2917 & -1.51 & 0,0235\\
	   	\hline
	\end{tabular}
	\caption{{\protect \small Energy shifts in meV, up to order  $\alpha^2$, when ${\tilde \theta=\alpha}$, for $n=1,2,3$.}}
	\label{tabla4}
\end{center}
\end{table}

\begin{figure}[H]
\centering
\includegraphics[width=0.7 \linewidth]{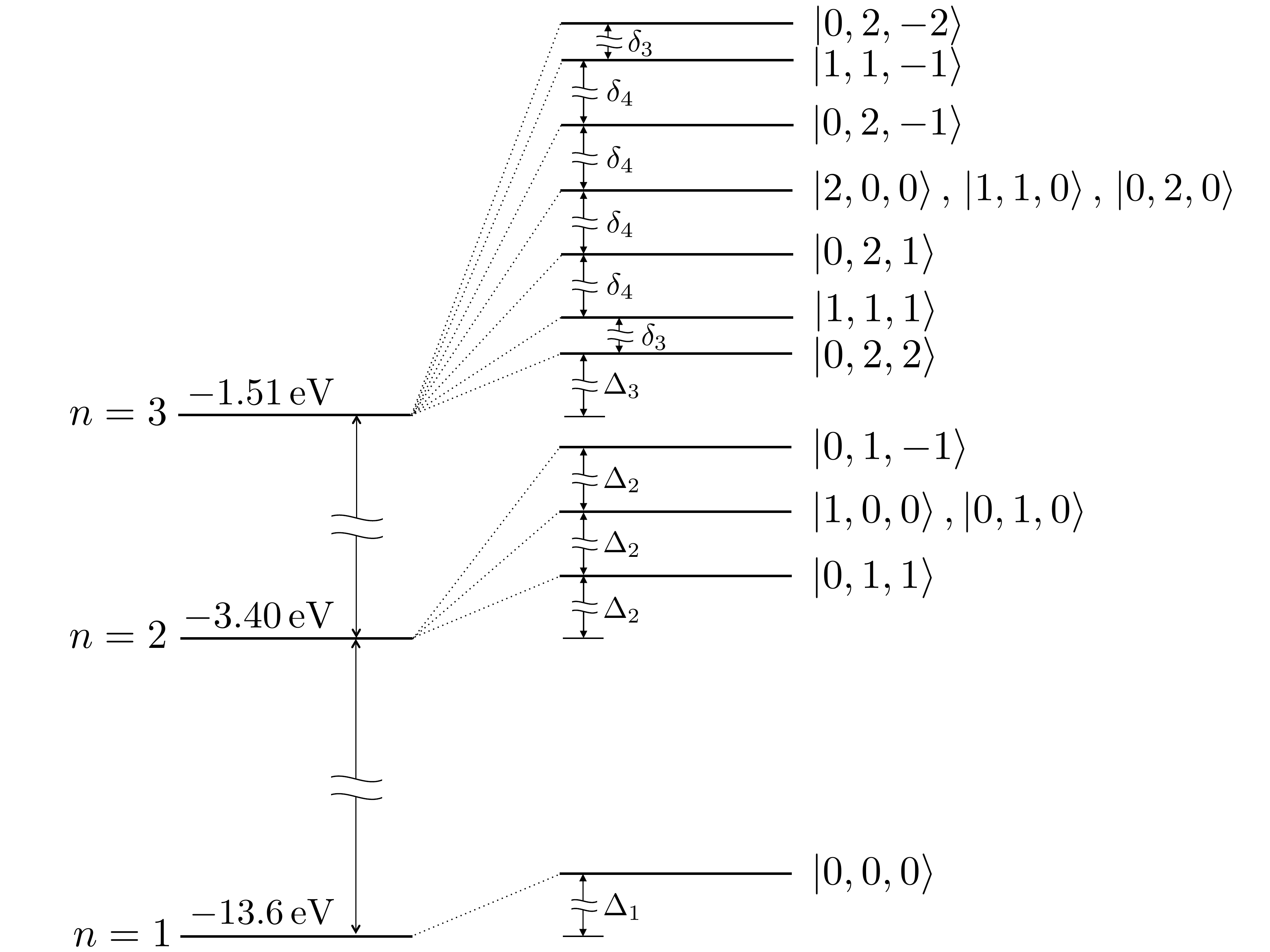}
\caption{{\protect \small Schematic of the splitting (not to scale) in the first three energy levels of the hydrogen atom in the presence of a $\theta$-vacuum. The states $|\nu,l, m \rangle$ denote the zeroth order contribution to the wave function. Here, $\Delta _{1} = 0.36$ meV, $\Delta _{2} = 0.0453$ meV, $\Delta _{3} = 0.0235$ meV, $\delta _{3} = 0.0033$ meV and $\delta _{4} \approx 2 \delta _{3}$.}}
\label{NivelesEnergia}
\end{figure}

\begin{figure}
\centering
\includegraphics[width=0.495\linewidth]{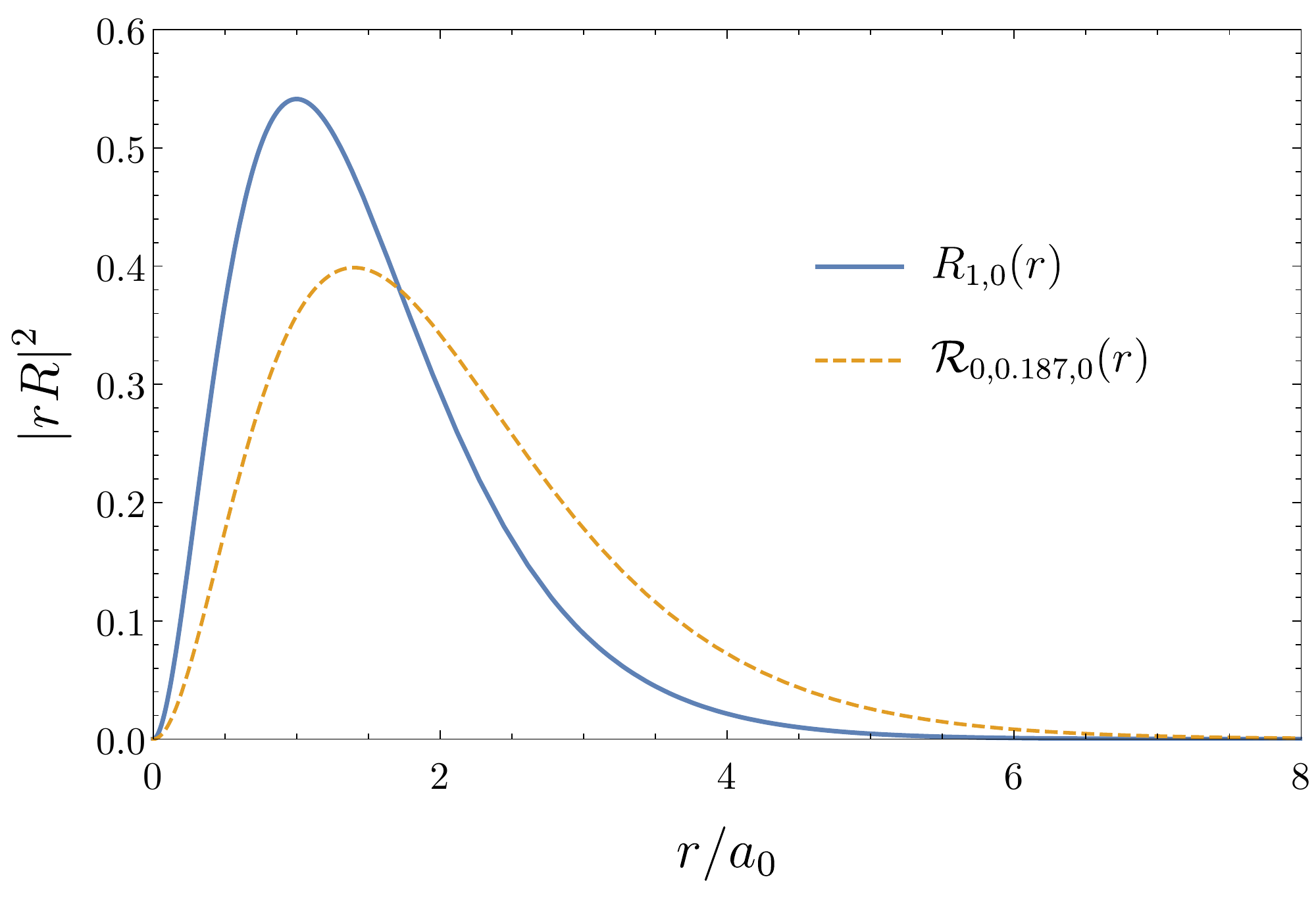}
\includegraphics[width=0.495\linewidth]{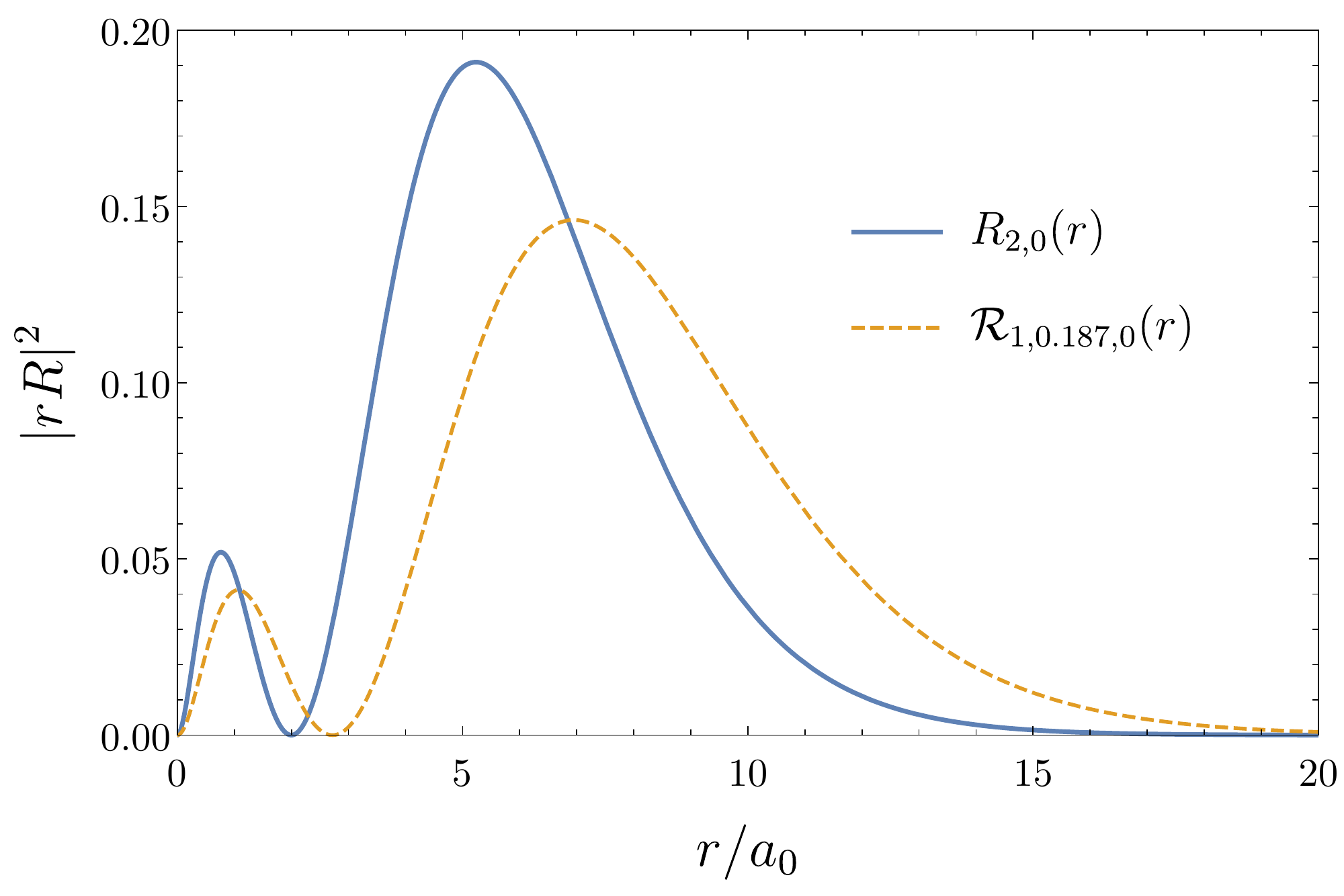}
\includegraphics[width=0.495\linewidth]{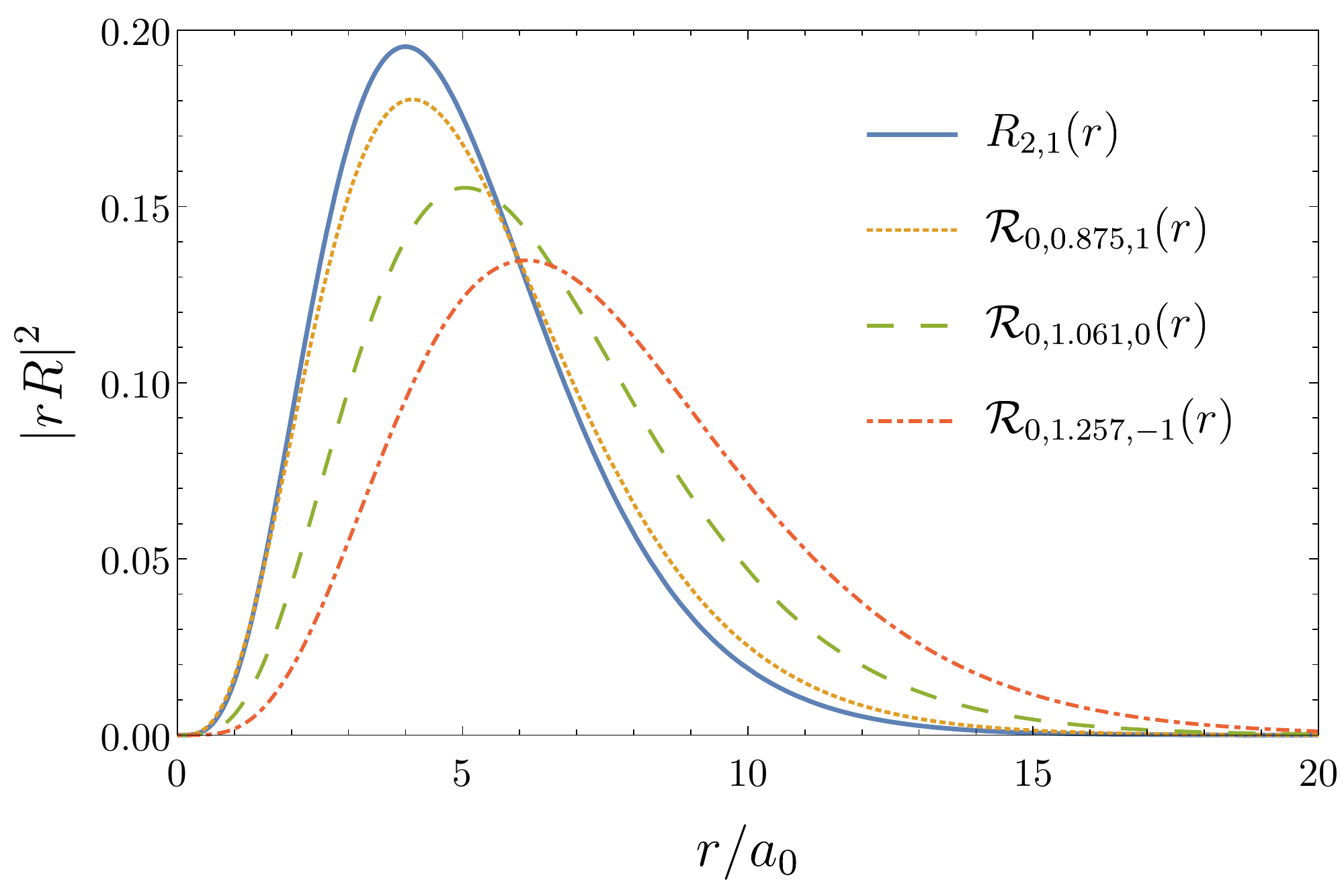}
\includegraphics[width=0.495\linewidth]{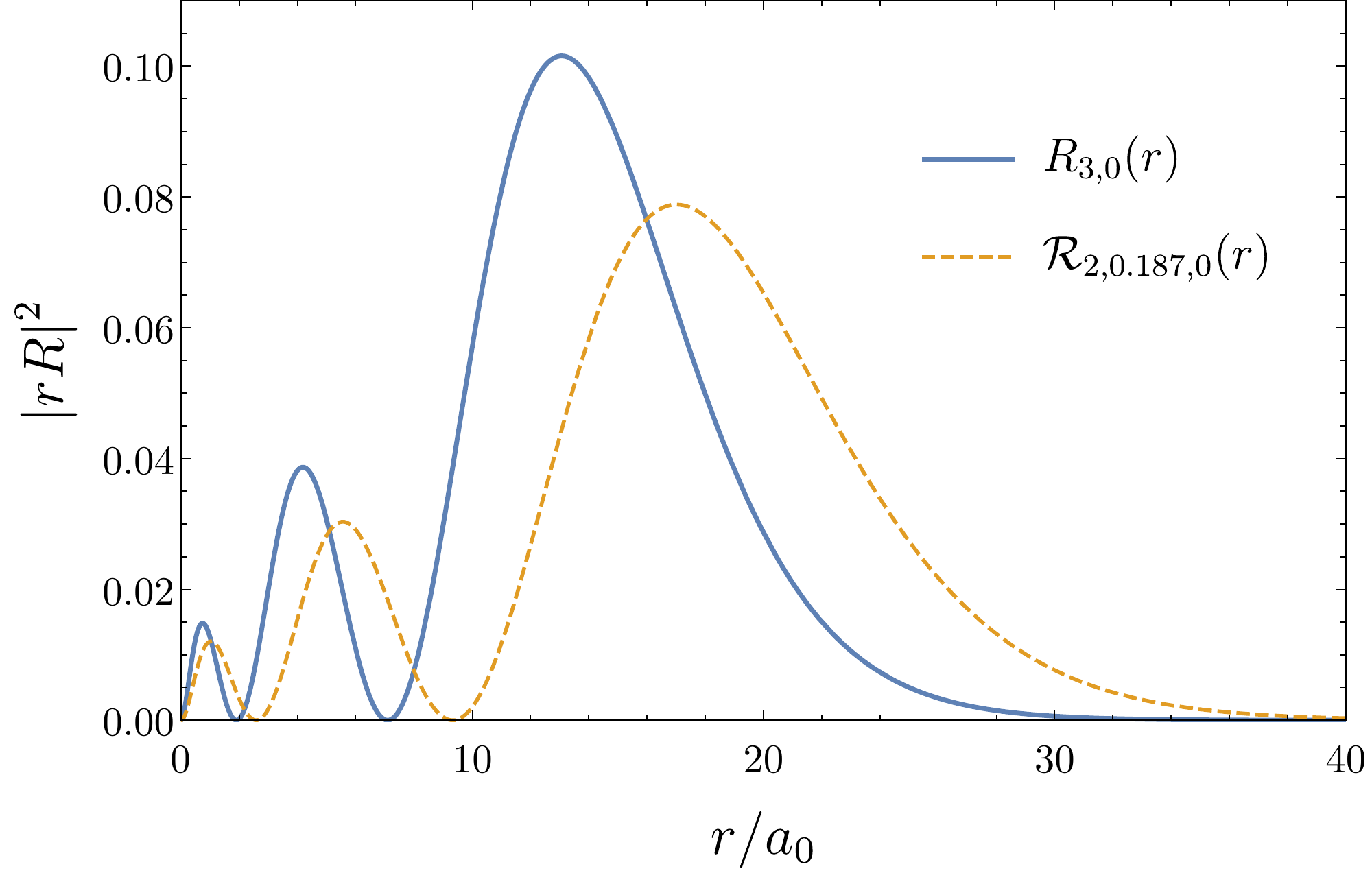}
\includegraphics[width=0.495\linewidth]{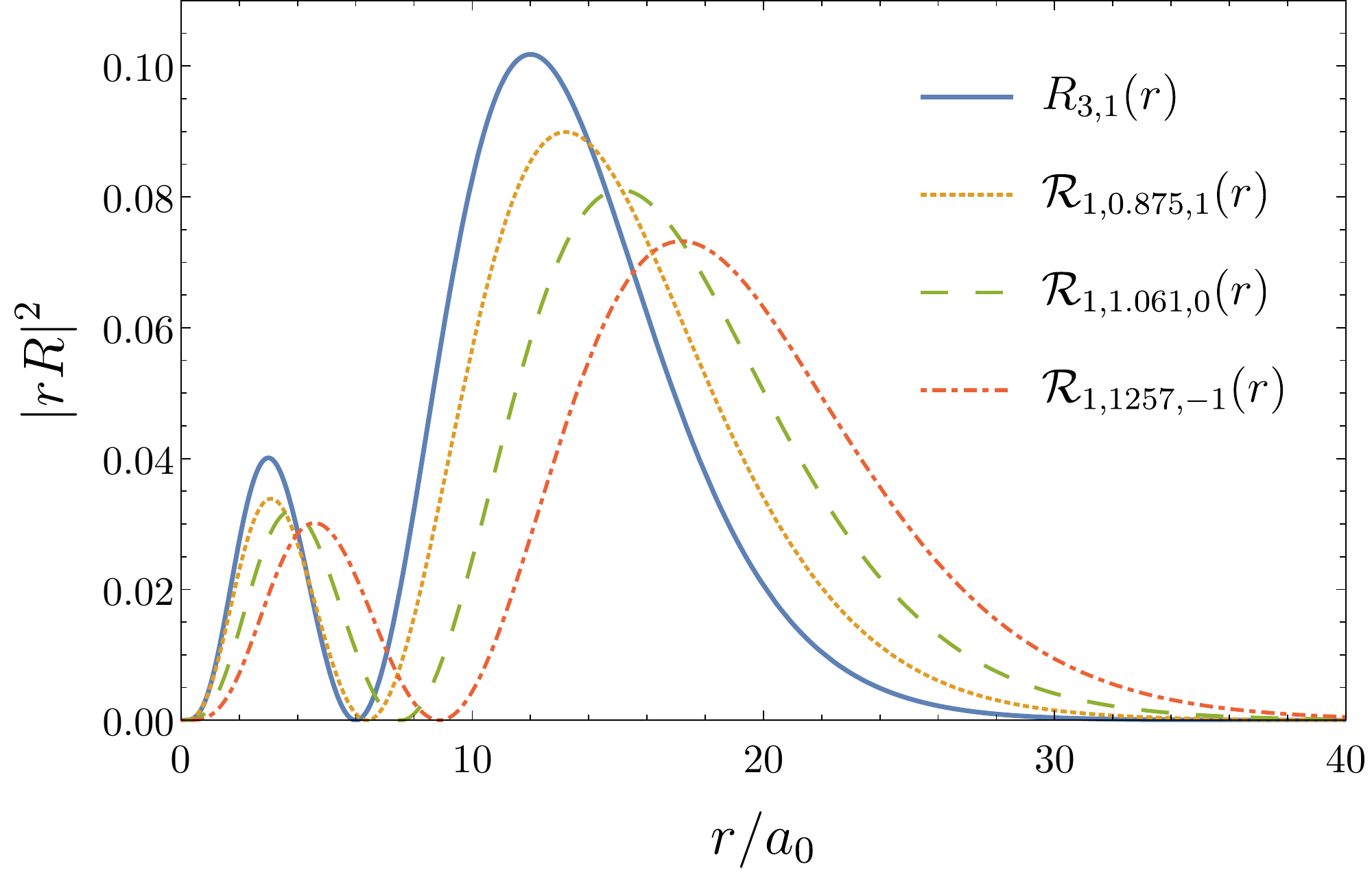}
\includegraphics[width=0.495\linewidth]{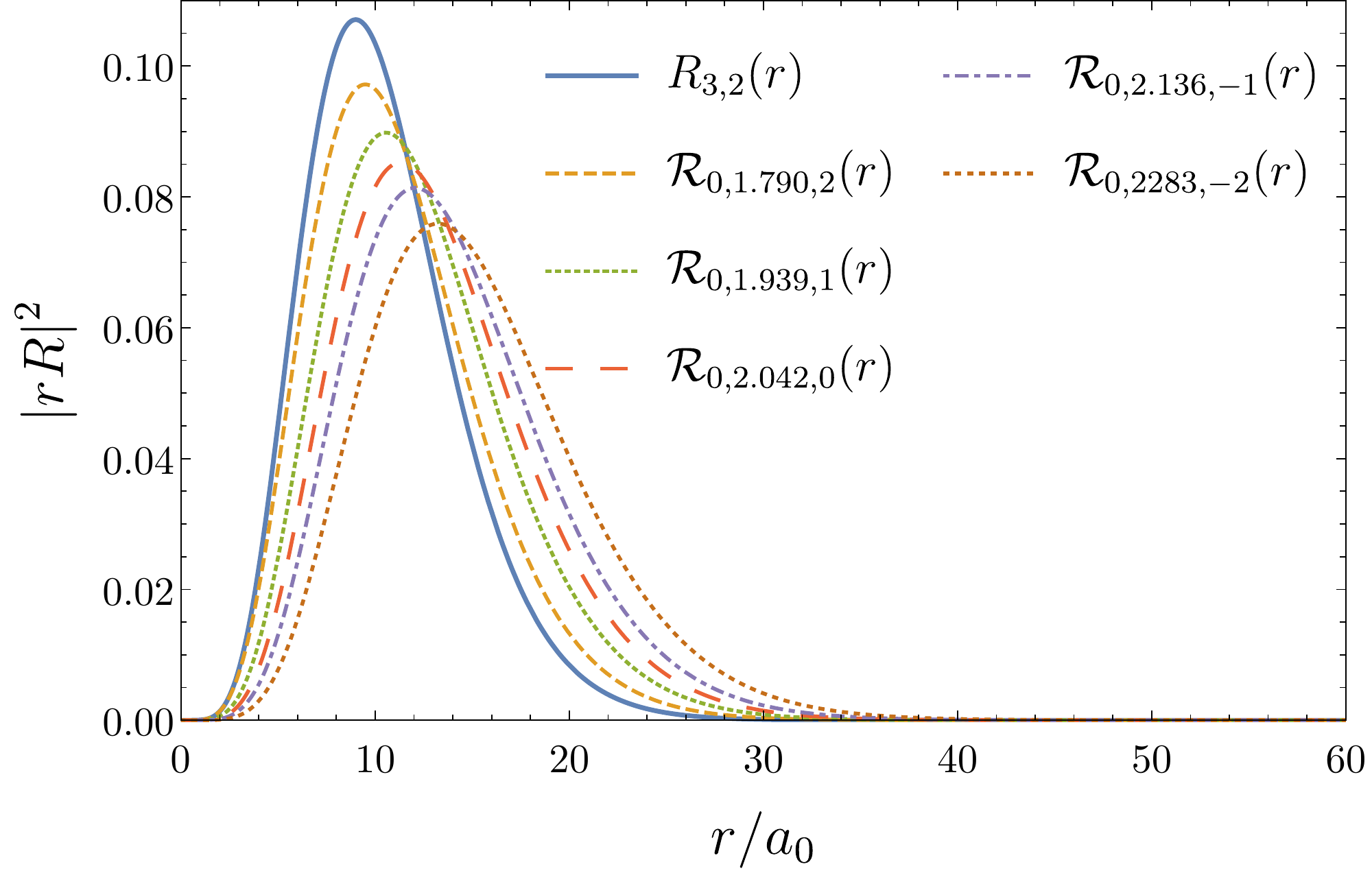}
\caption{{\protect \small Plots of the probability density $| r {\cal R}_{\nu, L, m}(r)|^2 $ together with the corresponding expressions for the hydrogen atom $|r R_{n,l}(r)|^2$. Here $a_0$ denotes the Bohr radius of the hydrogen atom and we have taken $a=0.2$.}}
\label{GraficaRadialA}
\end{figure}

\begin{table}
\begin{center}
	\begin{tabular}{||c|c|c||}
		\hline\hline
	    $l$ & $m$ & $L$ \\ 
		\hline \hline
	     0 & 0 & 0.1867\\
	     \hline
	     \multirow{3}{1cm}{\centering 1}  
	      & 0 & 1.0605\\
	      \cline{2-3}
	      &-1 & 1.2573\\
	      \cline{2-3}
	      & 1 & 0.8746\\
	     \hline
	     \multirow{3}{1cm}{\centering 2}
	      & 0 & 2.0415\\
	       \cline{2-3}
	      & -2 & 2.2832\\
	       \cline{2-3}
	      & -1 & 2.1362\\
	       \cline{2-3}
	      & 1  & 1.9392\\
	       \cline{2-3}
	      & 2  & 1.7910\\
	     \hline
	\end{tabular}
	\caption{{\protect \small Values of  $L$ for the choice  $a=0.2$ which are used in the plots appearing in Fig. \ref{GraficaRadialA}}}
	\label{tablaL}
\end{center}
\end{table}

\section{Calculation of the energy shifts using perturbation theory}

\label{PERT_CALC}

In this section we apply perturbation theory to obtain the corrections to the hydrogen atom spectra in front of the planar $\theta$-medium in the realistic case when $\tilde{\theta} \ll 1$. We start with the Hamiltonian operator (\ref{schrodinger1}):
\begin{align}
H &= - \frac{1}{2 \mu} \nabla ^{2} - \frac{e ^{2}}{r} + \frac{1}{2 \mu} \left[ \frac{2 \tilde{\theta}}{4 + \tilde{\theta} ^{2}} \frac{e ^{2}}{r ^{2}} \left\lbrace \frac{2}{1 + \vert \cos \vartheta \vert}  \right\rbrace \; i \frac{\partial}{\partial \varphi}\,\, +  \frac{4 \tilde{\theta} ^{2}}{(4 + \tilde{\theta} ^{2} ) ^{2}} \frac{e ^{4}}{r ^{2}} \left\lbrace\frac{2}{1 + \vert \cos \vartheta \vert} - 1 \right\rbrace  \right] + \frac{\tilde{\theta} ^{2}}{4 + \tilde{\theta} ^{2}} \frac{e ^{2}}{r} \notag \\ & \equiv  H _{0} + V(r, \vartheta) , \label{H+V}
\end{align}
where the first two terms in (\ref{H+V}), denoted by $H _{0}$,  correspond to the unperturbed hydrogen atom Hamiltonian with the zeroth order wave functions denoted by $|n,l,m\rangle$. The additional terms $V(r,\vartheta)$ are treated as a perturbation. The azimuthal symmetry of the problem allows to replace $i \frac{\partial}{\partial \varphi}$ by $- m$. Also, since we are dealing with a medium with  $\tilde{\theta}=\alpha$, we consider an expansion of $V(r,\vartheta)$ in powers of $\alpha$. Thus we work with the perturbation
\begin{align}
V(r,\vartheta) \equiv V _{1} + V _{2} + V _{3} =  - \frac{1}{2 \mu}\frac{\tilde{\theta} ^{2}}{2} \frac{1}{r ^{2}}\left\lbrace \frac{2}{1 + \vert \cos \vartheta \vert} \right\rbrace m\,\, + \frac{1}{2 \mu}\frac{\tilde{\theta} ^{4}}{4} \frac{1}{r ^{2}}\left\lbrace \frac{2}{1 + \vert \cos \vartheta \vert} - 1 \right\rbrace + \frac{\tilde{\theta} ^{3}}{4} \frac{1}{r} + \dots  .
\end{align}
The orders of magnitude of the above contributions in the perturbation are estimated as
\begin{align}
V _{1} \sim m \mu \alpha ^{4} , \qquad V _{2} \sim \mu \alpha ^{6} , \qquad V _{3} \sim \mu \alpha ^{4} .
\end{align}
That is to say, to the lowest order $\mu \alpha ^{4}$, which is within the scale of the fine-structure corrections, we consider only the terms $V _{1}$ and $V _{3}$, where we emphasize that $V _{1}$ contributes only for $m \neq 0$. The neglected term $V _{2}$ arises from the ${\bf A}^2$ contribution to the Hamiltonian. With the exception of the ground state, the remaining energy levels are degenerate in $l$ and $m$. Then, for the perturbation $V = V _{1} + V _{2}$ we need to calculate the matrix elements  $\langle n,l,m|V|n,l',m' \rangle$. Since $m$ is a good quantum number such matrix elements will be of the form  $\langle n,l,m|V|n,l',m'\rangle = \xi(n,l,m)\delta _{mm'}$ where  $\xi(n,l,m)$ is to be determined. 

The matrix element for $V _{3}$ is given by
\begin{align}
\langle n,l,m|V _{3}|n,l',m \rangle =\frac{\tilde{\theta} ^{3}}{4} \int _{0} ^{\infty} d r \,\, r ^{2} R _{n,l} (r) \frac{1}{r} R _{n,l'}(r) \int d \Omega Y ^{*m} _{l} (\Omega) Y ^{m}_{l'} (\Omega),
\end{align}
where the orthogonality of the spherical harmonics yields $l=l'$. Moreover, the well known result 
\begin{align}
\Bigg \langle \frac{1}{r} \Bigg \rangle _{n,l}= \int _{0} ^{\infty} d r \,\, r ^{2} R _{n,l}\frac{1}{r} R _{n,l} = \frac{\mu \alpha}{n ^{2}}, \label{1/r}
\end{align}
leads to 
\begin{equation}
\langle n,l,m| V _{3} |n,l',m \rangle = \frac{\mu \alpha ^{4}}{4 n ^{2}} \delta _{ll'}. \label{elementoV3}
\end{equation}
In the calculation of the matrix elements for $V _{1}$ we make use of the Pasternack-Sternheimer formula \cite{Pasternack,Cunningham},
\begin{equation}
\int _{0} ^{\infty} d r \,\, r ^{2} R _{n,l}(r)\frac{1}{r ^{s}} R _{n,l ^{\prime}}(r) = 0 \quad; \quad s = 2,3,\cdots,l-l^{\prime} + 1 \quad ; \quad l>l^{\prime}, \label{Pasternack}
\end{equation}
which implies that the only nonzero matrix elements for $s=2$ are those with $l=l'$. In this way, together with the expression 
\begin{align}
\Bigg \langle \frac{1}{r ^{2}} \Bigg \rangle _{n,l} = \int _{0} ^{\infty} dr \,\, r ^{2} R _{n,l} \frac{1}{r ^{2}} R _{n,l} = \frac{\mu ^{2} \alpha ^{2}}{n ^{3} (l+\frac{1}{2})}, \label{1/r2}
\end{align}
we have
\begin{align}
\langle n,l,m|V _{1} |n,l',m \rangle &= \delta_{l l'} \int d ^{3} x R _{n,l} (r) Y ^{*m} _{l}(\vartheta , \varphi) \left(\frac{- m \tilde{\theta} ^{2}}{4 \mu r ^{2}} \frac{2}{1 + \vert \cos \vartheta \vert} \right) R _{n,l'} Y ^{m} _{l'} (\vartheta,\varphi) , \nonumber \\
&= -2 m \pi \frac{\mu \alpha ^{4}}{n ^{3} \left(l + \frac{1}{2} \right)} \delta _{ll'} \int _{0}^{\frac{\pi}{2}} d \vartheta  \frac{\sin \vartheta \vert Y ^{m} _{l} (\vartheta,\varphi) \vert ^{2}}{1 + \cos\vartheta}, \label{elementoV1}
\end{align}
where the angular integral is calculated numerically in each case.

In this way, the nondiagonal terms of the matrix elements are zero and the energy shifts are given by 
\begin{align}
\Delta E _{n,l,m} = \langle n,l,m| (V _{1} + V _{3} ) | n,l,m \rangle,
\end{align}
even though we are dealing with a degenerate system.

The results are presented in the Table \ref{tabla5}.
\begin{table}
\begin{center}
	\begin{tabular}{||c|c|c|c|c||}
		\hline\hline
		$n$ & $l$ & $m$ & $\Delta E_{n,l,m}/(\mu \alpha^4)$ &$\Delta E_{n,l,m}$ (meV)\\ 
		\hline \hline
	     1& 0 & 0 & ${1}/{4}$ & 0,3620 \\ 
		\hline 
		\multirow{4}{1cm}{\centering 2} & 0 & 0 & $1/16$  &0,09050\\
		\cline{2-5}
		& \multirow{3}{1cm}{\centering 1} & -1  & $3/32$ &0,13580 \\ 
		\cline{3-5}
	    &  & 0   & $1/16$ & 0,09050 \\
	    \cline{3-5}
	    &  & 1   & $1/32$ &0,04526\\
	    \hline 
	    \multirow{9}{1cm}{\centering 3} & 0 & 0   & $1/36$ &0,04023 \\
	    \cline{2-5} 
	    & \multirow{3}{1cm}{\centering 1} & -1  & $1/27$ &0,05364\\ 
	   	\cline{3-5}
	   	&  & 0   & $1/36$ &0,04023\\
	    \cline{3-5}
	   	&  & 1    & $1/54$ &0,02682\\
        \cline{2-5}
	   	& \multirow{5}{1cm}{\centering 2} & -2   & $17/432$ &0,05670 \\
	   	\cline{3-5}
	   	&  & -1   & $7/216$ &0,04694 \\
	   	\cline{3-5}
	   	&  & 0   & $1/36$ &0,04023\\
	   	\cline{3-5}
	   	&  & 1  & $5/216$ &0,03352\\
	   	\cline{3-5}
	   	&  & 2  & $7/432$ &0,02347\\
	   	\hline
	\end{tabular}
	\caption{{\protect \small Corrections to the first energy levels of the hydrogen atom using first order perturbation theory.}}
	\label{tabla5}
\end{center}
\end{table}
The comparison of the perturbative results in Table \ref{tabla5} with those of the exact numerical calculation in Table \ref{tabla4} show a very good agreement. This also indicates that approximation to order  $\alpha^2$ in Eq. (\ref{EnergiaOrdenCuad}) is a very reasonable one in the case considered. 

\section{Summary and conclusions}
\label{SUMM}

The present  work relies on the idea that  atomic spectroscopy could play an important role in the study of the TME in axionic matter, as proposed in Refs. \cite{EPL} and \cite{PRA}.
We consider it as providing  an idealized setup where one can acquire a qualitative understanding of the orders of magnitude involved, together with a general  idea of the major modifications to be expected in the spectrum.} We have solved analytically the Schr\"{o}dinger equation for an hydrogen atom located at the interface between two $\theta$-media.  Interestingly enough, our setup can be viewed also as the solution of the motion of the electron in the magnetic field of a Pearl vortex \cite{Pearl,Flavio}. 

A $\theta$-medium is described by a particular  extension of electrodynamics
that consists of the Maxwell Lagrangian supplemented
by the coupling to the abelian Pontryagin density through a nondynamical  field $\theta$, restricted to the case where $\theta$ is piecewise constant in different regions of space separated by a common boundary. In this scenario the field equations in the bulk remain the standard Maxwell's equations, but the discontinuity of $\theta$ alters the behavior of the fields at the interface between media. Realistic versions of $\theta$-media are provided by topological insulators \cite{TI4}, where $\theta$ is quantized in odd integer values of $\pi$. The  sharp interface is  more than a useful approximation, being the result of an effective theory arising from the integration of the fermionic degrees of freedom describing the detailed microscopic structure of the material involved. Any coordinate dependent modification on the parameter $\theta$, designed to join continuously  both values across the interface, has to be interpreted as the appearance of another material besides the topological insulator. For example, a linear interpolation between the two values of $\theta$ across the interface would correspond  to  the introduction of a Weyl semimetal in the corresponding region.   What makes possible a smooth  transition between $\theta$-media is the breaking of time-reversal symmetry at the interface, which can be achieved by depositing a thin magnetic coating there, with a width of the order of $10^{-7}$cm. This magnetization $\mathbf{M}$ will generate an additional Zeeman effect on the electron, which we have not considered in calculating the splitting of the spectrum. Nevertheless, as reported in Ref.\cite{PRA}, this additional effect can be disentangled by measuring the energy shifts for different values of the magnetization and finally extrapolating to $\mathbf{M}=0$.

Here we have focused on the purely topological effects, and thus we considered both media with neither dielectric nor magnetic properties, i.e. $\varepsilon = \mu = 1$. It has been  previously shown that taking both media with the same dielectric constant allows for an enhancement of the topological  consequences of the TME effect, which are then not masked by  standard optical contributions \cite{PRA}. In this way, when an hydrogen atom is brought near to the interface between $\theta$-media, the modified field equations (\ref{MAXMOD}) predict additional electromagnetic interactions which in turn will affect the quantum-mechanical motion of the atomic electron.

Here we have considered the simple case in which the $\theta$-interface is the plane $z = 0$, and we derived the Schr\"{o}dinger equation for an hydrogen atom nearby. In order to deal with an analytically tractable problem we make the further approximation of localizing the nucleus, which we assume to be at rest, exactly at the $\theta$-interface. In a more realistic case, this could serve as a zeroth-order approximation to the case of a Rydberg hydrogen atom close to the interface between media. Under these assumptions, it is possible to solve the Schr\"{o}dinger equation exactly and analytically by using separation of variables in spherical coordinates. The system has azimuthal symmetry with respect to the $z$-axis, and thus the magnetic quantum number $m$, which is the eigenvalue of $L _{z}$, remains as a good quantum number. The translational symmetry along the $z$ direction is broken due to the presence of the interface, and consequently the resulting polar angular equation introduces a deformation $L$ of the standard angular momentum quantum number $l$. The solutions are given in terms of Gauss hypergeometic functions in Eqs. (\ref{SolPx,m>0})-(\ref{SolPx,m<0}) in the region $0 < \vartheta < \pi/2$. The quantization of $L$ is obtained numerically by demanding such solutions to be even or odd under parity transformations, according to the conditions (\ref{Solpar}) and (\ref{Solimpar}). We find that two parameters, $\tilde{e}$ (which describes the effective Coulomb interaction) and $a$ (which takes care of the magnetoelectric effect), control the shape of the polar angular solution. Here we take  $\tilde \theta$ arbitrary, and consequently $\tilde e$ and $a$,  in order to amplify the effects  of the magnetoelectric polarizability upon the energy shifts and wave functions. Some illustrative  results are listed in Tables \ref{tabla2a}  and \ref{tabla1b} for the choices $a = 1$ and $a _{\mbox{\scriptsize m}} = 1.82 \times 10 ^{-3}$, respectively. We observe that in the latter case, the smallness of $a _{\mbox{\scriptsize m}}$ is sufficient to practically wipe out the difference between $L$ and $l$. This is even more dramatic in the case of a realistic $\theta$-medium with $\tilde{\theta} = \alpha \, \,  (a=1.33 \times 10^{-5})$, which is shown in Table {\ref{tabla1}. Some examples of the deformed angular functions ${\cal P}^m_L$ together with the standard ones $P^m_{l}$ are plotted in Fig \ref{Figura} for the case $a=1$.

An additional deformation $\tilde \ell$ of the angular momentum quantum number, given in Eq. (\ref{ell}), allows to write the solutions of the radial equation (\ref{ec.radial3}) in terms of the standard solutions for the hydrogen atom, yielding the corresponding energy eigenvalues (\ref{Energia}). The numerical results for the  energy shifts of the levels corresponding to $n=1,2,3$ to order ${\tilde \theta}^2$, defined in Eq. (\ref{DELTAE}), are shown in Table \ref{tabla4} for $\tilde \theta=\alpha$. They range between $0.4$ and  $0.02$ meV,  being smaller than fine structure corrections which are of the order of meV. These corrections to the energy levels are also calculated to first order in perturbation theory  with the results in Table \ref{tabla5} showing a very good agreement with those of Table \ref{tabla4}. The breaking of the degeneracy in the hydrogen energy levels to order $\alpha^2 $ induced by the magnetoelectric effect is made explicit in those tables and it is further illustrated in Fig. \ref{NivelesEnergia}. Let us observe  that for each value $n$, the lines corresponding to $l, m=0$ remain degenerate because the correction is of order $\alpha^4$ in these cases. The probability density $\rho_{\nu,L,m}=  |r {\cal R}_{\nu,L,m}|^2$ for $\nu=0,1,2$ is plotted in Fig. \ref{GraficaRadialA} and compared with the  analogous probability density $\vert r R _{nl} \vert ^{2}$ of the hydrogen atom, which is independent of $m$, showing the breaking of the degeneracy in $m$ within each subspace labeled by $l$. Finally we calculate the  energy shifts $\Delta E_{n,l,m}$ in the cases $n=1,2,3$, using first order perturbation theory when $\tilde \theta=\alpha$. Even though each subspace $n$ is degenerate, with the exception of the ground state, the corrections to the energy levels are given only by the diagonal matrix elements of the perturbation.  This is a consequence of the orthogonality of the spherical harmonics together with the use of the Pasternak-Sternheimer formula \cite{Pasternack,Cunningham}, as explained in section \ref{PERT_CALC}. The results are presented in Table \ref{tabla5} and they show a very good agreement with those in Table \ref{tabla4} calculated in the numerical approximation to the exact solution.
 It should be clear that the physics of our problem is completely different from that of an electron moving in the field of a dyon. Nevertheless, the similarity of the vector potentials (\ref{POTFIN}) and (\ref{POTMON}) may lead to some confusion and  demands a careful explanation of the differences involved, which we have presented in Section \ref{COMP_MON}.

\acknowledgments   This work is supported in part by Project
No. IN104815 from Direcci\'{o}n General Asuntos del Personal Acad\'{e}mico
(Universidad Nacional Aut\'{o}noma de M\'{e}xico) and CONACyT (M\'{e}xico),
Project No. 237503. AMR was supported by the CONACyT postdoctoral  grant \# 234774.

\section*{Author contribution statement}

DAB, AM-R and LFU  contributed  to the  the theoretical analysis, to the
calculations and  to the writing of the manuscript.

\end{document}